\documentclass[lettersize,journal]{IEEEtran}
\usepackage{amsmath,amsfonts}
\usepackage{algorithmic}
\usepackage{algorithm}
\usepackage{array}
\usepackage[caption=false,font=normalsize,labelfont=sf,textfont=sf]{subfig}
\usepackage{textcomp}
\usepackage{stfloats}
\usepackage{url}
\usepackage{verbatim}
\usepackage{graphicx}
\usepackage{cite}
\usepackage{amssymb}
\usepackage{multirow}
\usepackage{booktabs}
\usepackage[table,xcdraw]{xcolor}
\usepackage{natbib}
\usepackage{amsmath}
\setcitestyle{numbers,square}
\usepackage[colorlinks,bookmarksopen,bookmarksnumbered,citecolor=blue, linkcolor=red, urlcolor=black]{hyperref}
\usepackage{cite}

\begin{document}

\title{Feature Compression for Cloud-Edge Multimodal 3D Object Detection}

\author{Chongzhen Tian, Zhengxin Li, Hui Yuan, Raouf Hamzaoui, Liquan Shen, and Sam Kwong,
\thanks{This work was supported in part by the National Natural Science Foundation of China under Grants 62222110 and 62172259, the High-end Foreign Experts Recruitment Plan of Chinese Ministry of Science and Technology under Grant G2023150003L, the Taishan Scholar Project of Shandong Province (tsqn202103001), the Natural Science Foundation of Shandong Province under Grant ZR2022ZD38. (Chongzhen Tian and Zhengxin Li contribute equally. Corresponding author: Hui Yuan)}
\thanks{Chongzhen Tian, Zhenxin Li and Hui Yuan are with the School of Control Science and Engineering, Shandong University, Jinan 250061, China (e-mail: cztian@mail.sdu.edu.cn; lizx@mail.sdu.edu.cn; huiyuan@sdu.edu.cn).

Raouf Hamzaoui is with the School of Engineering and Sustainable Development, De Montfort University, LE1 9BH Leicester, U.K. (e-mail: rhamzaoui@dmu.ac.uk).

Liquan Shen is with the Key Laboratory of Specialty Fiber Optics and Optical Access Networks, Joint International Research Laboratory of Specialty Fiber Optics and Advanced Communication, Shanghai University, Shanghai 200444, China (e-mail: jsslq@163.com).

Sam Kwong is with the School of Data Science, Lingnan University, Hong Kong (e-mail: samkwong@ln.edu.hk).}}

\markboth{Journal of \LaTeX\ Class Files,~Vol.~14, No.~8, August~2021}%
{Shell \MakeLowercase{\textit{et al.}}: A Sample Article Using IEEEtran.cls for IEEE Journals}


\maketitle
\begin{abstract}
Machine vision systems, which can efficiently manage extensive visual perception tasks, are becoming increasingly popular in industrial production and daily life. Due to the challenge of simultaneously obtaining accurate depth and texture information with a single sensor, multimodal data captured by cameras and LiDAR is commonly used to enhance performance. Additionally, cloud-edge cooperation has emerged as a novel computing approach to improve user experience and ensure data security in machine vision systems. This paper proposes a pioneering solution to address the feature compression problem in multimodal 3D object detection. Given a sparse tensor-based object detection network at the edge device, we introduce two modes to accommodate different application requirements: Transmission-Friendly Feature Compression (T-FFC) and Accuracy-Friendly Feature Compression (A-FFC). In T-FFC mode, only the output of the last layer of the network's backbone is transmitted from the edge device. The received feature is processed at the cloud device through a channel expansion module and two spatial upsampling modules to generate multi-scale features. In A-FFC mode, we expand upon the T-FFC mode by transmitting two additional types of features. These added features enable the cloud device to generate more accurate multi-scale features. Experimental results on the KITTI dataset using the VirConv-L detection network showed that T-FFC was able to compress the features by a factor of 6061 with less than a 3\% reduction in detection performance. On the other hand, A-FFC compressed the features by a factor of about 901 with almost no degradation in detection performance. We also designed optional residual extraction and 3D object reconstruction modules to facilitate the reconstruction of detected objects. The reconstructed objects effectively reflected the shape, occlusion, and details of the original objects.

\end{abstract}
\vspace{-0.1cm} 
\begin{IEEEkeywords}
Feature compression, multimodal, 3D object detection, sparse tensor, feature coding for machines.
\end{IEEEkeywords}
\vspace{-0.3cm} 
\section{Introduction}
\label{Sec-I}
\IEEEPARstart{M}{achine} vision models are increasingly replacing human vision systems for repetitive tasks, offering higher accuracy and faster decision-making \cite{ref1}. Accurate recognition of the surrounding scene is fundamental to achieving high performance in machine vision. To this end, multimodal data, often represented as RGB-X, has been widely adopted in machine vision systems. Depending on the scene characteristics and the perceptual task, X can represent depth, thermal, polarimetric, event information, or other data \cite{ref2}. Among these data types, the combination of image and point cloud data has found extensive applications in autonomous driving, unmanned aerial vehicle navigation, environmental monitoring, and virtual reality \cite{ref3}. While multimodal data provides richer information, it also significantly increases data volume, posing challenges for transmission and storage. This is particularly critical in cloud-edge collaborative computing scenarios with limited bandwidth, such as ocean object detection, traffic flow analysis, and robot rescue operations. In these contexts, the efficiency of multimodal data compression directly impacts the performance of machine vision systems. In this paper, we focus on the task of multimodal 3D object detection and explore efficient feature compression methods.

LiDAR captures depth information by transmitting pulsed laser beams and detecting the scattered light characteristics of objects. It is known for its high precision, extensive range, and strong anti-interference capabilities. However, due to the monochromatic nature of the laser, it cannot capture color and texture information. Additionally, LiDAR-based detection models experience significant performance degradation when objects are far from the sensor, as point density decreases with distance. In contrast, cameras can capture detailed texture and color information, and the pixel density remains consistent regardless of distance. However, image sensors lack effective depth perception. Therefore, combining LiDAR and camera data is a promising approach to achieve more accurate object detection results \cite{ref4}. Recent research has focused on fusing point cloud data from LiDAR with image data by converting images into pseudo or virtual points \cite{ref5}. This fusion strategy not only addresses the limitations of single-modal data but also simplifies the input format for detection models. Building on this fusion framework, we propose a compression method that can be applied to both point cloud-based single-modal detection methods and point cloud and image-based multimodal detection methods, offering broad applications.

In the cloud-edge collaboration mode, the edge device is used to capture scene information and transmit it to the cloud device. The cloud device receives the transmitted data and conducts a downstream analysis. Two main approaches are typically used: Compression Then Analysis (CTA) \cite{ref6} and Analysis Then Compression (ATC) \cite{ref6}. In CTA, the edge device compresses the captured data and transmits it to the cloud. The cloud device receives the compressed data and reconstructs it before processing it. Therefore, the compression operations conducted at the edge stage are task-independent which may lead to a limited compression ratio. Moreover, the whole downstream task network is implemented in the cloud device, which causes excessive pressure on cloud computing and slows down the response time when the cloud device needs to serve multiple edge devices. 

ATC can be split into result compression and feature compression. The main difference between them is that the former compresses the results of perception tasks while the latter compresses the intermediate features. In the result compression approach, the captured data is directly sent to the downstream task network at the edge device, and the perception results are compressed and transmitted. The cloud device only needs to decompress the received data to obtain the perception results. The advantage of this approach is that only a small amount of data needs to be transmitted. However, direct transmission of the results is very sensitive to channel noise. This method also poses a risk to privacy as the transmitted data directly reflects the final results. Moreover, the use of cloud-edge computing resources is unbalanced since the whole downstream task is conducted in the edge device while the computing resources of cloud devices are not fully exploited. 

In the feature compression approach, the downstream task network is divided into two parts. The edge device extracts downstream task-related features from the captured data, compresses the data, and transmits the compressed data. The cloud device reconstructs the received data and then feeds the reconstructed data into the second part of the machine vision perception network to obtain the task results. Compared with the raw data compression approach, higher compression ratios can be achieved by exploiting prior information about perception tasks. Compared with the results compression approach, the feature compression approach typically requires transmitting more data but offers better privacy protection and stronger resistance to channel noise. This is because the features must pass through the second part of the network to generate human-interpretable results. Therefore, even if the features are intercepted, it is difficult to decipher their meaning without the corresponding network model. Moreover, the network has a certain level of robustness \cite{ref7} and can usually generate outputs with reduced error when the input data is partially changed. More importantly, collaborative intelligence in the third approach can achieve a better allocation of computing resources between the cloud and edge devices. However, the feature volume is usually greater than the volume of the raw input data. This is because the multi-channel expression of the features leads to greater sparsity. Hence, feature compression requires a higher compression efficiency. Considering the above factors, the feature compression approach may be the optimal choice. Therefore, we focus on designing an efficient feature compression method to reduce data transmission pressure and support cloud-edge collaboration.

In summary, while multimodal data enhances perception accuracy, it also increases the pressure on data transmission. Moreover, relying only on the cloud device or edge device for downstream perception inevitably leads to many issues, such as data transmission delay, inefficient use of computing resources, threats to privacy, and sensitivity to noise. On the other hand, transmitting features poses greater challenges for the design of the compression scheme. To the best of our knowledge, no feature compression method for multimodal 3D object detection currently exists. This paper addresses this gap as follows.

\textbf{1. Feature Compression in Multimodal 3D Object Detection:} We explore feature compression in multimodal 3D object detection and propose two modes to adapt to varying requirements. Our encoder and decoder can integrate with any sparse tensor-based detection method. Additionally, the compressed features can be used to reconstruct detected objects using residual information. This method significantly enhances machine perception efficiency.

\textbf{2. T-FFC Mode:} In the T-FFC mode, we introduce a masked transformer enhancement module, a coarse-grained compression module, and a fine-grained compression module to enhance object areas and efficiently compress the feature from the backbone's last layer. Furthermore, a spatial upsampling module and a channel expansion module enable efficient feature reconstruction with minimal performance reduction at the cloud device. 

\textbf{3. A-FFC Mode:} In the A-FFC mode, we incorporate two additional branches to transmit extra features alongside the basic T-FFC features. The channel compression of these additional features mirrors that of the basic feature, and spatial compression of these newly transmitted features is achieved using the basic feature and a backward mapping operation. These new features provide more precise geometric information, resulting in better detection performance.

\textbf{4. Residual Injection-based Object Reconstruction:} To fully exploit transmitted features and provide an intuitive object representation at the cloud device, we propose a residual injection-based object reconstruction method. The reconstruction leverages the existing point cloud upsampling method, designed around features commonly transmitted in both modes and additional residual information. The inclusion of residual information enhances reconstruction quality. 

The remainder of this paper is organized as follows. Section \ref{Sec-II} reviews the related works. Section \ref{Sec-III} states the research problem. Section \ref{Sec-IV} introduces the proposed methods. Section \ref{Sec-IV} presents our experimental results. Finally, Section \ref{Sec-VI} draws conclusions and suggests future work.
\vspace{-0.3cm} 
\section{Related Works}
\label{Sec-II}
\subsection{3D Object Detection Methods}
3D object detection aims to recognize the objects of interest by drawing an oriented 3D bounding box and assigning a label, and plays a key role in collision avoidance, path planning, and motion prediction \cite{ref14}. Depending on the modality of the input data, 3D object detection methods can be classified into three categories: image-based methods, LiDAR-based methods, and multimodal methods. Compared with LiDAR, camera is cheap and flexible. However, the primary problem that needs to be solved for 3D object detection in images is the lack of depth information. Philion \textit{et al}. \cite{ref15} used the object detection loss to supervise object detection and depth estimation. Li \textit{et al}. \cite{ref16} provided explicit depth supervision to improve depth estimation performance. Chen \textit{et al}. \cite{ref17} proposed a binocular image-based 3D object detection method. Huang \textit{et al}. \cite{ref18} used the view transformer to convert multi-view image features to Bird’s-Eye View (BEV) for object prediction. However, binocular or multi-view systems usually require more hardware and software for calibration and processing and are greatly affected by illumination changes, thus limiting their applications.

Point cloud-based methods generally outperform image-based methods for 3D object detection. This is because point clouds can provide accurate depth information. From the perspective of representation granularity of point clouds, point cloud-based detection methods can be divided into three categories: point-based methods \cite{ref12}, voxel-based methods \cite{ref9}, and voxel-point-based methods \cite{ref24}. It has been shown that voxel-based object detection methods can achieve similar results to point-based detection methods while reducing computational complexity \cite{ref8}. Thus, voxel-based 3D object detection methods have obtained widespread attention in recent years. However, point cloud-based detection methods typically suffer from performance degradation when detecting distant objects due to the sparse sample density. 

The complementary advantages between point clouds and images encourage many researchers to study the 3D object detection problem with multimodal input data. From the perspective of the multimodal data fusion stage, the existing detection methods can be divided into three categories \cite{ref27}: early fusion \cite{ref28}, late fusion \cite{ref30}, and cascade fusion \cite{ref31}. The early fusion methods first conduct multimodal data fusion and then feed the fusion data to the detection model. This type of method can compensate for the lack of single-modal representation at the source and the object detection of fused data can be conducted in the existing single-modal object detection model. The late fusion methods first conduct object detection on a single modal data and then fuse the detection results to improve the detection accuracy. The computational complexity of the late fusion methods is usually higher than that of the early fusion methods. The cascade fusion methods first carry out the initial screening of the target area based on one input modal data to narrow the detection range of the other input modal data. The initial screening results greatly affect the final detection performance. Based on the above analysis, early fusion methods are lightweight and easy to implement. The mainstream approach converts the image into a virtual point cloud and then combines it with the captured or true point cloud to achieve a more accurate scene perception. Finally, the combined point clouds are fed into a 3D object detection network to make predictions. In this paper, we choose a lightweight but effective detection model (VirConv-L \cite{ref9}) as the detection baseline and compress the extracted features.
\vspace{-0.2cm} 
\subsection{Data Compression for Human Vision}
Compared with image and video compression, point cloud compression is more challenging as the data is sparse and unordered. MPEG set up the MPEG-3D group in 2017 to research the standardization of point cloud compression. Up to now, MPEG-3D has proposed Geometry-based Point Cloud Compression (G-PCC) and Video-based Point Cloud Compression (V-PCC) schemes. Besides MPEG-3D, the Audio Video Coding Standard Workshop of China (AVS) also proposed a corresponding Point Cloud Exploration Model (PCEM). In addition to these standards, many effective point cloud compression schemes have been developed. For example, Zhang \textit{et al}. \cite{ref45} introduced quad-tree and binary-tree partitions to further improve the geometric coding efficiency. Wang \textit{et al}. \cite{ref46} proposed deep learning-based compression methods to obtain better compression performance. Although these human vision system-based point cloud and image compression methods have achieved great success in recent years, they cannot be directly used for data compression for machine vision tasks. This is because the optimization target of these compression methods is to improve the similarity between the original data and the reconstructed data while reducing the data size \cite{ref48}. For machine perception tasks, the compressed data generated in this way may contain redundant information \cite{ref70}. Therefore, it is highly desirable to design efficient compression methods for machine vision systems that do not significantly degrade task-related information.
\vspace{-0.2cm} 
\subsection{Data Compression for Machine Vision}
Compression methods for the human visual system need to consider fidelity while compression methods for machine vision models should pay more attention to the characteristics of machine vision tasks \cite{ref71}. Generally, machine vision system-oriented compression methods can be divided into two categories: raw input-based compression methods and feature-based compression methods. Raw input-based compression methods can be further divided into three categories \cite{ref49}: side information-guided methods \cite{ref50}, machine vision-constrained methods \cite{ref51}, and semantic information preservation-based methods \cite{ref52}. For example, Huang \textit{et al}. \cite{ref50} designed a machine-vision-based distortion measurement model to evaluate the quality of task-related features and proved that using the proposed model to guide the bit allocation process can save 28.17\% bitrate without affecting the accuracy of visual analysis tasks. Le \textit{et al}. \cite{ref51} proposed an end-to-end training model using machine task-related loss and rate loss for efficient image compression. Yang \textit{et al}. \cite{ref52} proposed the semantic information preservation-based image compression method which uses a preprocessor to extract task-related semantic information and adopts the newly designed semantic distillation loss to optimize the compression model. However, the above methods implicitly combine the compression algorithm with the downstream task characteristics, which may lead to a limited compression ratio. 

Recently, many researchers have combined compression algorithms with downstream task characteristics. For example, Chen \textit{et al}. \cite{ref53} proposed the first feature compression method that uses HEVC as the compression tool and designed evaluation metrics for feature compression. Eshratifar \textit{et al}. \cite{ref55} proposed BottleNet, which uses a shallow and deep model respectively at the edge device and cloud device for compression-aware training. Shao \textit{et al}. \cite{ref56} present BottleNet++ to improve the compression ability of BottleNet \cite{ref55} by modeling the channel as a non-trainable layer. In general, the above-mentioned feature compression methods usually obtain a higher compression ratio than raw data compression methods. Nevertheless, these methods cannot be directly applied to machine vision-oriented point cloud compression tasks since the features of point clouds are disordered and sparse. In the task of point cloud compression, we need to compress both geometry and attribute information. 

Research on machine vision system-oriented point cloud compression is relatively new, with only a few related works. For example, Lu \textit{et al}. \cite{ref57} showed that conducting moderate compression does not affect task performance. Liu \textit{et al}. \cite{ref58} proposed the first compression method for both human and machine vision tasks. Zhao \textit{et al}. \cite{ref59} introduced a semantic prior representation to guide the compression process for LiDAR point clouds. However, these schemes compress points rather than features, which limits the compression ratio. Machine vision task-based feature compression for LiDAR point clouds remains an open challenge. This paper addresses this challenge by aiming to eliminate redundancy from point cloud features while preserving machine perception performance.

\vspace{-0.3cm}  
\section{Problem Statement}
\label{Sec-III}
\subsection{Definition of Different Compression Strategies}
\subsubsection{Information Entropy Lossless Compression} Given the input data \emph{D}, since there is spatial redundancy (caused by the correlation between adjacent pixels/points in the image or point cloud), time redundancy (caused by the correlation between different frames in image or point cloud sequence), and spectral redundancy (caused by the correlation of different color planes or spectral bands) in raw data, the Information Entropy LossLess compression representation form $\emph{D}_{\mathrm{IELL}}$ exists. Generally, the redundancy can be represented as 
\vspace{-0.1cm} 
\begin{equation}
\begin{aligned}
\label{gongshi1}
\delta_{\mathrm{IELL}} &= \mathbb{P} \left\{ \left\| \emph{D}- \emph{D}_{\mathrm{IELL}} \right\| \right\} \\
                       &= \mathop{\underbrace{\mathbb{P} \left\{ \Delta \emph{D}_{\mathrm{Spa}}  \right\}}}\limits_{\delta_{\mathrm{Spa}}}
                       +\mathop{\underbrace{\mathbb{P} \left\{ \Delta \emph{D}_{\mathrm{T}}  \right\}}}\limits_{\delta_{\mathrm{T}}}
                       +\mathop{\underbrace{\mathbb{P} \left\{ \Delta \emph{D}_{\mathrm{Spe}}  \right\}}}\limits_{\delta_{\mathrm{Spe}}},
\end{aligned}
\vspace{-0.1cm} 
\end{equation}
where  $\delta_{\mathrm{IELL}}$ is the redundancy calculated through information entropy lossless compression strategy (which can be approximately represented as a combination of spatial redundancy $\delta_{\mathrm{Spa}}$, time redundancy $\delta_{\mathrm{T}}$ and spectral redundancy $\delta_{\mathrm{Spe}}$), $\mathbb{P} \left\{ \cdot \right\}$ denotes a specific pooling operation, $\Delta \emph{D}_{\mathrm{Spa}} $ denotes the data difference introduced by spatial redundancy (the same definition for $\Delta \emph{D}_{\mathrm{T}}$ and $\Delta \emph{D}_{\mathrm{Spe}}$). The optimization objective of this compression strategy can be denoted as
\begin{equation}
\begin{aligned}
\label{gongshi2}
\mathop{\mathrm{minimize}}\limits_{\theta_{\mathrm{IELL}}} \ &\mathbb{R} \left\{ \mathbb{C}_{\mathrm{IELL}}  \left\{\emph{D},\theta_{\mathrm{IELL}}  \right\} \right\} \\
s.t. \ \mathbb{H} \left\{\emph{D} \right\} &= \mathbb{H} \left\{ \mathbb{C}_{\mathrm{IELL}} \left\{\emph{D}, \theta_{\mathrm{IELL}}\right\} \right\}, 
\end{aligned}
\end{equation}
where $\mathbb{R} \left\{ \cdot \right\} $ denotes the bitrate calculation operation,   $\mathbb{C}_{\mathrm{IELL}} $ denotes the information entropy lossless compression algorithm with parameter $\theta_{\mathrm{IELL}} $, and $\mathbb{H} \left\{ \cdot \right\}$ denotes the information entropy calculation operation. 

\subsubsection{Downstream Task Perception Lossless Compression}When considering the needs of downstream tasks (e.g., human/machine perception), information entropy lossless compression might not be the optimal strategy. This is because the subject of the downstream tasks (e.g., human vision system and machine vision algorithm) can only perceive distortion beyond a certain range. Research has shown that there is just noticeable distortion in both human \cite{ref65} and machine \cite{ref66} perception applications. Therefore, researchers have turned to designing perceptual lossless compression strategies to achieve higher compression ratios. Let $\emph{D}_{\mathrm{DTLL}}$ denote the reconstructed data obtained through the downstream task perception lossless compression algorithm. Compared with $\emph{D}_{\mathrm{DTLL}}$, the redundancy in \emph{D} is 
\begin{equation}
\begin{aligned}
\label{gongshi3}
\delta_{\mathrm{DTLL}}\! &=\! \mathbb{P}\! \left\{ \left\| \emph{D}\!-\! \emph{D}_{\mathrm{DTLL}} \right\| \right\} \\
                       &=\! \mathop{\underbrace{\mathbb{P}\! \left\{ \Delta \emph{D}_{\mathrm{Spa}}  \right\}}}\limits_{\delta_{\mathrm{Spa}}}
                       \!+\!\mathop{\underbrace{\mathbb{P}\! \left\{ \Delta \emph{D}_{\mathrm{T}}  \right\}}}\limits_{\delta_{\mathrm{T}}}
                       \!+\!\mathop{\underbrace{\mathbb{P}\! \left\{ \Delta \emph{D}_{\mathrm{Spe}}  \right\}}}\limits_{\delta_{\mathrm{Spe}}}
                       \!+\!\mathop{\underbrace{\mathbb{P}\! \left\{ \Delta \emph{D}_{\mathrm{P}}  \right\}}}\limits_{\delta_{\mathrm{P}}},
\end{aligned}
\end{equation}
where  $\delta_{\mathrm{P}}$  denotes the perceptual redundancy that the subject of downstream tasks can tolerate. The optimization objective of this compression algorithm can be expressed as
\begin{equation}
\begin{split}
\label{gongshi4}
\mathop{\mathrm{minimize}}\limits_{\theta_{\mathrm{DTLL}}} \ \mathbb{R} \left\{ \mathbb{C}_{\mathrm{DTLL}} \left\{ \emph{D} , \theta_{\mathrm{DTLL}} \right\} \right\} \ \ \ \ \ \\
s.t. \ \mathbb{A} \left\{ \mathbb{C}_{\mathrm{DTLL}} \left\{ \emph{D}, \theta_{\mathrm{DTLL}} \right\} \right\} - \mathbb{A} \left\{ \emph{D} \right\} \geq 0,
\end{split}
\end{equation}
where $\mathbb{C}_{\mathrm{DTLL}} \left\{ \cdot \right\}$ denotes the downstream task perception lossless compression algorithm with parameter $\theta_{\mathrm{DTLL}}$, and $\mathbb{A} \left\{ \cdot \right\}$ denotes the downstream task accuracy evaluation method. Generally speaking, the perceptual coding algorithm can effectively remove perceptual redundancy and thus improve performance on downstream tasks. 
\subsubsection{Downstream Task Perception Lossy Compression}To transmit and store data with a huge amount of volume in a limited channel, researchers turn to research downstream task perception lossy compression methods and introduce rate-distortion optimization technique. Let $\emph{D}_{\mathrm{DTLY}}$ denotes the reconstructed data obtained through Downstream Task perception LossY compression algorithm. The redundancy in \emph{D} can be expressed as
\vspace{-0.1cm} 
\begin{equation}
\begin{aligned}
\label{gongshi5}
\delta_{\mathrm{DTLY}} &= \mathbb{P} \left\{ \left\| \emph{D}- \emph{D}_{\mathrm{DTLY}} \right\| \right\} = \mathop{\underbrace{\mathbb{P} \left\{ \Delta \emph{D}_{\mathrm{Spa}}  \right\}}}\limits_{\delta_{\mathrm{Spa}}}+\mathop{\underbrace{\mathbb{P} \left\{ \Delta \emph{D}_{\mathrm{T}}\right\}}}\limits_{\delta_{\mathrm{T}}} 
                       \\&+\mathop{\underbrace{\mathbb{P} \left\{ \Delta \emph{D}_{\mathrm{Spe}}  \right\}}}\limits_{\delta_{\mathrm{Spe}}}
                       +\mathop{\underbrace{\mathbb{P} \left\{ \Delta \emph{D}_{\mathrm{P}}  \right\}}}\limits_{\delta_{\mathrm{P}}}
                       +\mathop{\underbrace{\mathbb{P} \left\{ \Delta \emph{D}_{\mathrm{PL}}  \right\}}}\limits_{\delta_{\mathrm{PL}}},
\end{aligned}
\vspace{-0.1cm} 
\end{equation}
where $\delta_{\mathrm{PL}}$ denotes the data that causes a perceptual loss in the downstream task. The larger the  $\delta_{\mathrm{PL}}$, the smaller the data volume. The final cost of a downstream task lossy compression algorithm consists of two components: a bitrate cost and a performance cost, denoted as
\vspace{-0.1cm} 
\begin{equation}
\label{gongshi6}
\emph{L}\!=\!\mathbb{R}\!\left\{\mathbb{C}_{\mathrm{DTLY}}\!\left\{\emph{D},\theta_{\mathrm{DTLY}}\right\}\!\right\}\!+\!\lambda\!\cdot\!\left\{\mathbb{A}\!\left\{\emph{D}\right\}\!-\!\mathbb{A}\!\left\{\emph{D}_{\mathrm{DTLY}}\right\}\!\right\}\!,
\end{equation}
where $\mathbb{C}_{\mathrm{DTLY}}\left\{\cdot\right\}$ denotes the downstream task lossy compression algorithm with parameter $\theta_{\mathrm{DTLY}}$, and $\lambda$ denotes the parameter used for tuning the degree of accuracy loss.
\vspace{-0.3cm} 
\subsection{Differences Between Data Coding for Human and Machine}
Generally, human-oriented compression methods aim to establish a compromise between compressed data volume and users’ subjective experience while machine-oriented compression methods aim to establish a compromise between compressed data volume and machine tasks’ performance.  
\begin{figure*}
    \centering
    \includegraphics[width=1\linewidth]{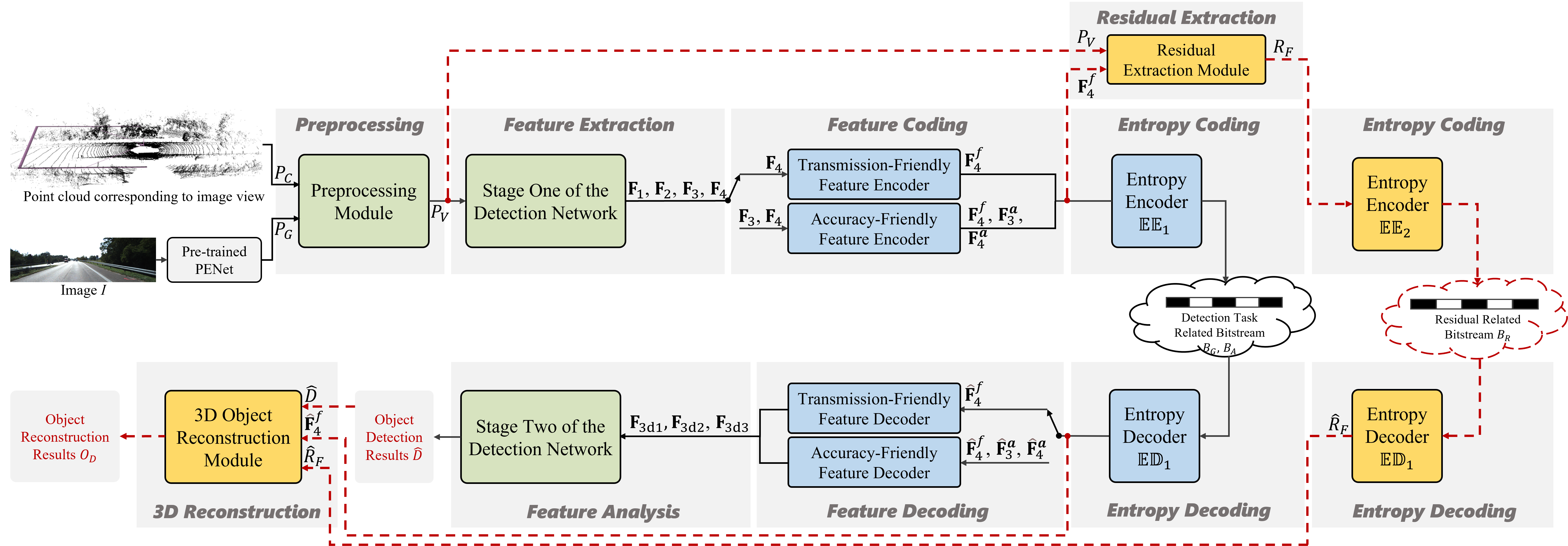}
    \caption{Overall framework of the proposed method. Reconstruction task-related operations are displayed with red dotted lines to distinguish them from detection task-related operations.}
    \label{figure2}
\vspace{-0.6cm} 
\end{figure*}
As for human perception scenes, the challenges of designing efficient compression algorithms come from two aspects. On the one hand, we need to design an accurate objective quality assessment model to simulate subjective quality assessment process since subjective quality assessment is expensive and time-consuming. However, the human vision system is complex and highly nonlinear, which makes simple quality metrics (e.g., MSE and MS-SSIM) cannot obtain optimal results. Later, more and more quality metrics are introduced to obtain more inconsistent cognition with subjective perception \cite{ref67}. On the other hand, we need to design efficient compression schemes to precisely find out data redundancy. In the beginning, many traditional coding algorithms are designed to continuously improve the compression ratios. Recently, more and more researchers attempt to design deep learning-based modules to replace partial or all operations of traditional algorithms. The results show that the introduction of deep learning technology can bring in higher compression ratios.

As for machine perception scenes, the downstream task performance evaluation can directly use the existing machine task performance measurement metrics. However, designing an efficient machine task-friendly compression algorithm is challenging. The initial methods try to conduct compression operation on the raw data by designing machine task-guided bit allocation scheme \cite{ref50}, bringing in machine task-related loss \cite{ref51}, or bringing in semantic information restriction \cite{ref52} to achieve machine-oriented compression. However, the disadvantage of this kind of methods lies in that we need to conduct complete data analysis operation on the cloud device, which is not suitable for cloud-edge collaboration scenario. Moreover, this combination of compression methods and downstream task characteristics is inadequate, which can only produce a suboptimal compression ratio. The other mainstream machine-oriented compression methods try to compress the features extracted by the downstream task network rather than the raw input data. The advantage of this kind of methods is twofold. First, the downstream task network only extracts the task-specific feature in the input data and thus can obtain a higher compression ratio. Second, the feature compression scheme provides the possibility for separately conducting the feature extraction and feature analysis operations, which can make full use of cloud and edge computing resources. The loss function of feature compression can be represented as
\vspace{-0.1cm} 
\begin{equation}
\begin{aligned}
\label{gongshi7}
\emph{L} = \mathbb{R} \bigl\{ \mathbb{C}_{\mathrm{DTLY}}^\mathrm{F} \bigl\{ \mathbb{N}_{\mathrm{DT}}^{1} \bigl\{ \emph{D} , \alpha_{\mathrm{DT}}^{1*} \bigl\}, \theta_{\mathrm{DTLY}}^\mathrm{F} \bigl\} \bigl\} + \lambda\cdot \ \ \ \ \ \ \\ \bigl\{ \mathbb{A} \bigl\{ \mathbb{N}_{\mathrm{DT}}^{2} \bigl\{ \mathbb{N}_{\mathrm{DT}}^{1} \bigl\{ \emph{D} , \alpha_{\mathrm{DT}}^{1} \bigl\}\!, \alpha_{\mathrm{DT}}^2 \bigl\}\!\bigl\} -\mathbb{A}\bigl\{\mathbb{N}_{\mathrm{DT}}^{2} \bigl\{\hat{\textbf{F}}, \alpha_{\mathrm{DT}}^{2*}  \bigl\}\!\bigl\}\!\bigl\},
\end{aligned}
\vspace{-0.1cm} 
\end{equation}
where $\mathbb{C}_{\mathrm{DTLY}}^\mathrm{F}$ denotes the feature-oriented downstream task lossy compression algorithm with parameter $\theta_{\mathrm{DTLY}}^\mathrm{F}$, $\mathbb{N}_{\mathrm{DT}}^{1}\left\{ \cdot\right\}$ denotes the part one (i.e., feature extraction part) of the downstream network with parameter $\alpha_{\mathrm{DT}}^1$ or $\alpha_{\mathrm{DT}}^{1*}$,  $\mathbb{N}_{\mathrm{DT}}^{2}\left\{ \cdot\right\}$ denotes the part two (i.e., feature analysis part) of the downstream network with parameter $\alpha_{\mathrm{DT}}^2$ or $\alpha_{\mathrm{DT}}^{2*}$. Considering that the downstream task model might be joint optimization with the compression models, we use * to distinguish different network parameters. $\hat{\textbf{F}}$ denotes the reconstructed feature.

However, the compression of features might be more challenging since features are sparser than raw input data. Moreover, the existing downstream task perception networks tend to extract multi-scale features to provide a more comprehensive perception of the input data. Obviously, compressing multi-scale features is more difficult than compressing single-scale raw data. In this paper, we focus on the problem of feature compression and provide two optional modes: A-FFC and T-FFC. In the A-FFC mode, we transmit multi-scale features at the edge device and the cloud device can obtain near-lossless detection performance. Besides, we also provide an optional object reconstruction branch. Considering that the reconstruction task needs more information than the detection task, we transmit an additional residual to increase the amount of information entropy that the cloud device can receive.
\vspace{-0.2cm} 
\section{Proposed Method}
\label{Sec-IV}
\vspace{-0.1cm} 
As shown in Fig. \ref{figure2}, the image $I$ is first fed to the pre-trained PENet \cite{ref60} to generate the virtual point cloud $P_G$. Then, $P_G$ and the captured point cloud $P_C$ are sent to the preprocessing module to obtain the voxel representation $P_V$. Next, the VirConv-L \cite{ref9} feature extraction module extracts detection-related features ($\textbf{F}_\mathrm{1}$, $\textbf{F}_\mathrm{2}$, $\textbf{F}_\mathrm{3}$, $\textbf{F}_\mathrm{4}$). In the original VirConv-L, $\textbf{F}_\mathrm{3}$ and $\textbf{F}_\mathrm{4}$ are sent to stage two of the detection network to conduct feature analysis and obtain the detection results. Specifically, $\textbf{F}_\mathrm{4}$ is first sent to the Region Proposal Network (RPN) to generate region proposals. Then, the region proposals, $\textbf{F}_\mathrm{3}$, and $\textbf{F}_\mathrm{4}$ are sent to the detection head to obtain the final prediction results. Considering that $\textbf{F}_\mathrm{4}$ is used in two different modules in the feature analysis part, we represent the inputs to feature analysis part as $\textbf{F}_\mathrm{3}$, $\textbf{F}_\mathrm{4}^{r}$, and $\textbf{F}_\mathrm{4}^{d}$. $\textbf{F}_\mathrm{4}^{r}$ should pay more attention to semantic enhancement to assist in accurate object location while $\textbf{F}_\mathrm{3}$ and $\textbf{F}_\mathrm{4}^{d}$ should pay more attention to detail retention to assist in fine box regression. Based on this analysis, we provide two optional modes (Transmission-Friendly Feature Encoder and Accuracy-Friendly Feature Encoder) in the feature coding part to accommodate different requirements. The former conducts feature coding operations only on the most important feature $\textbf{F}_\mathrm{4}$ to obtain $\textbf{F}_\mathrm{4}^{f}$, while the latter conducts feature coding operations on $\textbf{F}_\mathrm{3}$ and $\textbf{F}_\mathrm{4}$ to obtain $\textbf{F}_\mathrm{4}^{f}$, $\textbf{F}_\mathrm{3}^{a}$, $\textbf{F}_\mathrm{4}^{a}$. Then, the compressed sparse tensor is fed into entropy encoder $\mathbb{EE}_1$ to generate the detection task-related bitstreams $\emph{B}_{G}$ and $\emph{B}_{A}$. On the decoder side, the entropy decoder $\mathbb{ED}_1$ is used to decode the bitstream to features. Next, the feature decoding part is used for feature reconstruction. Corresponding to the feature coding, it also contains two optional modes: Transmission-Friendly Feature Decoder and Accuracy-Friendly Feature Decoder. The former attempts to generate three kinds of features ($\textbf{F}_\mathrm{3d1}$, $\textbf{F}_\mathrm{3d2}$, and $\textbf{F}_\mathrm{3d3}$) from $\hat{\textbf{F}}_\mathrm{4}^{f}$ to replace $\textbf{F}_\mathrm{4}^{r}$, $\textbf{F}_\mathrm{3}$ and $\textbf{F}_\mathrm{4}^{d}$. while the latter attempts to generate corresponding features from $\hat{\textbf{F}}_\mathrm{4}^{f}$, $\hat{\textbf{F}}_\mathrm{3}^{a}$, $\hat{\textbf{F}}_\mathrm{4}^{a}$. In addition, we design optional Residual Extraction (RE) and 3D Object Reconstruction (3DOR) modules to reconstruct the objects. The entropy encoder $\mathbb{EE}_2$ and entropy decoder $\mathbb{ED}_2$ are used to encode and decode the reconstruction task-related residual.
\subsection{Preprocessing}
In this module, $P_C$ and $P_G$ are combined and converted into a voxel representation denoted as $P_V=\left\{ \textbf{G}_V, \textbf{A}_V\right\}$. $\textbf{G}_V \in \mathbb{R}^{N\times 3}$ denotes the center point coordinates of the non-empty voxels, $\textbf{A}_V \in \mathbb{R}^{N\times 8}$ denotes the average attribute information of points contained in the voxels, $\emph{N}$ denotes the number of non-empty voxels. Specifically, the attribute information for each point includes the position coordinates $\emph{x}$, $\emph{y}$, $\emph{z}$, reflectance $\emph{r}$, color information $\emph{R}$, $\emph{G}$, $\emph{B}$, and flag $\emph{f}$. For the point in $P_C$, $\emph{f}$ is set to 2, and $\emph{R}$,  $\emph{G}$, and $\emph{B}$ are set to 0. For the point in $P_G$, $\emph{r}$ is set to 0, $\emph{f}$ is set to 1, and $\emph{R}$, $\emph{G}$, and $\emph{B}$ values are taken from the color values of the corresponding pixel in $I$. Moreover, if the number of points in the current voxel is greater than 5, we use random sampling to remove the excess points. If the number of points in the current voxel is less than 5, the missing attribute information is filled with zeros.
\begin{figure*}
    \centering
    \setlength{\abovecaptionskip}{0cm}
    \includegraphics[width=1\linewidth]{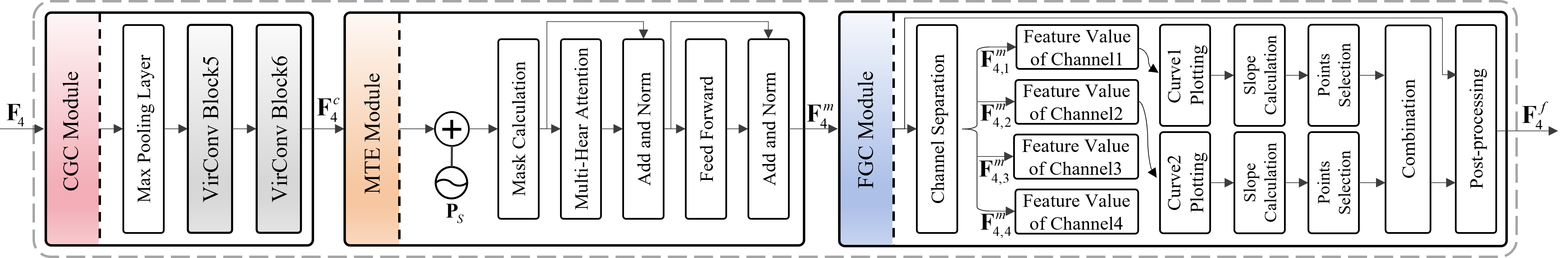}
    \caption{Transmission-friendly feature encoder.}
    \label{figure4}
\vspace{-0.6cm} 
\end{figure*}
\vspace{-0.1cm} 
\subsection{Feature Extraction}
The first stage of the 3D object detection network consists of four VirConv blocks. Each VirConv block comprises a stochastic voxel discard layer (which is used for discarding nearby redundant voxels), a 3D sparse convolution layer (which is used for restraining the noise caused by inaccurate depth estimation), and a noise-resistant submanifold convolution layer (which is used for down-sampling feature maps). The down-sampling scale of these blocks is set to 1, 2, 4 and 8 and the number of output channels of these blocks are set to 16, 32, 64 and 64. The outputs of VirConv blocks are denoted as $\textbf{F}_\mathrm{1}$, $\textbf{F}_\mathrm{2}$, $\textbf{F}_\mathrm{3}$ and $\textbf{F}_\mathrm{4}$, respectively. The first half of channels of $\left\{\textbf{F}_\emph{i} \left( \cdot \right) \mid \emph{i}=1,2,3,4 \right\}$ are obtained by 3D convolution operations, and the second half of channels of $\left\{\textbf{F}_\emph{i} \left( \cdot \right) \mid \emph{i}=1,2,3,4 \right\}$ are obtained by 2D convolution operations. Taking $\textbf{F}_\mathrm{4}$ as an example, the calculation process can be formulated as
\begin{equation}
\label{gongshi8}
\textbf{F}_\mathrm{4} = \mathcal{C}_\mathrm{4}\left(\mathcal{C}_\mathrm{3}\left(\mathcal{C}_\mathrm{2}\left(\mathcal{C}_\mathrm{1}(P_V)\right)\right)\right),
\end{equation}
where $\left\{ \mathcal{C}_i (\cdot) \mid i=1,2,3,4 \right\}$ denotes the $i$-th VirConv block. 
\subsection{Feature Coding}
\subsubsection{Transmission-Friendly Feature Encoder}
Fig. \ref{figure4} illustrates the pipeline of the transmission-friendly feature encoder, which consists of a coarse-grained compression (CGC) module, a masked transformer enhancement (MTE) module, and a fine-grained compression (FGC) module. A detailed description of these modules is given below.

(a) CGC Module: The CGC module consists of a max pooling layer and two VirConv blocks (VirConv Block5 and VirConv Block6). The former applies uniform spatial downsampling, and the latter is responsible for channel compression. The above operations can be formulated as
\begin{equation}
\label{gongshi9}
\textbf{F}_\mathrm{4}^c = \mathcal{C}_\mathrm{6} \left( \mathcal{C}_\mathrm{5} \left( \mathcal{M} \left( \textbf{F}_\mathrm{4} \right) \right) \right),
\end{equation}
where $\textbf{F}_\mathrm{4}^c$ denotes the output of the CGC module, and $\mathcal{M}\left( \cdot \right)$ denotes maxpooling. The maxpooling kernel size is set to 2×2×2. The number of output channels of $\mathcal{C}_\mathrm{5}$ and $\mathcal{C}_\mathrm{6}$ are set to 16 and 4, respectively.

(b) MTE Module: In the MTE module, we use masked attention calculation \cite{ref12} to achieve better performance with less training time than global attention calculation. Specifically, the sinusoidal position embedding $\textbf{P}_{S}$ is adopted to retain position information. Then, the masked transformer encoder $\mathcal{T}_{M}\left( \cdot \right)$ is used for feature enhancement. The above processes can be formulated as
\begin{equation}
\label{gongshi10}
\textbf{F}_\mathrm{4}^m = \mathcal{T}_{M}\left(\textbf{F}_{4}^{c}+\textbf{P}_{S}\right),
\end{equation}
where $\textbf{F}_\mathrm{4}^m$ denotes the output of the MTE module. In our experiment, the number of layers in the transformer encoder is set to 4, and the number of multi-heads is set to 4.

\begin{figure*}
    \centering
    \setlength{\abovecaptionskip}{0cm}
    \includegraphics[width=1\linewidth]{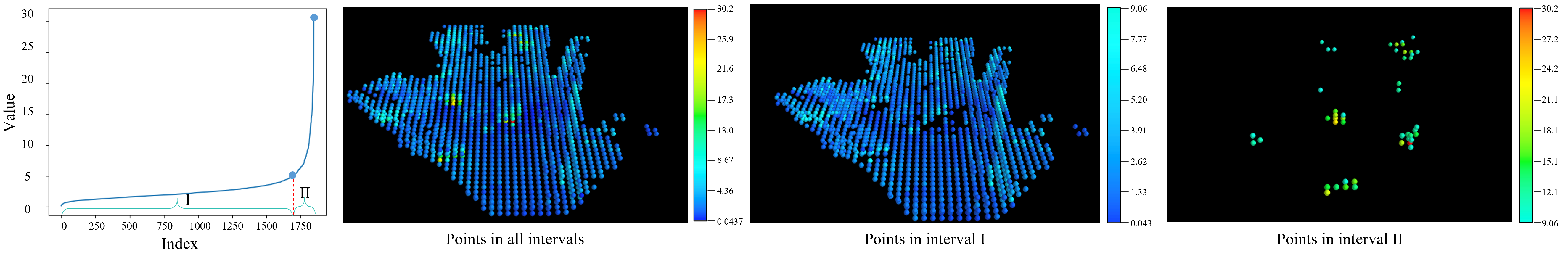}
    \caption{Slope-guided point selection.}
    \label{figure5}
\vspace{-0.6cm} 
\end{figure*}

(c) FGC Module: Although the maxpooling operation is conducted in the CGC module across the entire 3D space, there is still a lot of redundant spatial information. Specifically, the non-target area still exists in the scene but it is useless for downstream tasks. Therefore, we conduct fine-grained feature compression to further increase the compression ratio. Specifically, we first conduct channel separation on $\textbf{F}_\mathrm{4}^m$ to obtain $\textbf{F}_\mathrm{4,1}^m$, $\textbf{F}_\mathrm{4,2}^m$, $\textbf{F}_\mathrm{4,3}^m$, and $\textbf{F}_\mathrm{4,4}^m$. Then, all points in  $\textbf{F}_\mathrm{4}^m$, which can be denoted as $S = \left\{\textbf{P}_i \mid \emph{i}=1,2,...,N_4^m \right\}$, are sorted in ascending order according to the feature value in $\textbf{F}_\mathrm{4,1}^m$ and $\textbf{F}_\mathrm{4,2}^m$. The sequences of sorted points are denoted as $S_1 = \left\{\textbf{P}_{\mathrm{1,}i} \mid \emph{i}=1,2,...,N_4^m \right\}$ and $S_2 = \left\{\textbf{P}_{\mathrm{2,}i} \mid \emph{i}=1,2,...,N_4^m \right\}$, respectively. Taking the index $\emph{i}$ as the X-axis and the corresponding feature value in $\textbf{F}_\mathrm{4,1}^m$ as the Y-axis, we draw this curve in Fig. \ref{figure5}. This curve can be divided into two intervals since there is an obvious slope change at the right end of the curve. Points in the interval I belong to non-target area while points in the interval II belong to target area. Therefore, removing the points in interval I can suppress background noise and achieve efficient geometric compression. In the implementation, we first calculate the slope at each point of the curve. Then, we set a slope threshold $\emph{k}_\mathrm{TH}$ to select the points belonging to the object area. The selected points according to $\textbf{F}_\mathrm{4,1}^m$ are denoted as $S_1^s = \left\{\textbf{P}_{\mathrm{1,}i}^s \mid \emph{i}=1,2,...,N_{1}^s \right\}$, where $N_{1}^s$ denotes the number of points selected by $\textbf{F}_\mathrm{4,1}^m$.
\begin{figure}[]
    \centering
    \setlength{\abovecaptionskip}{0cm}
    \includegraphics[width=1\linewidth]{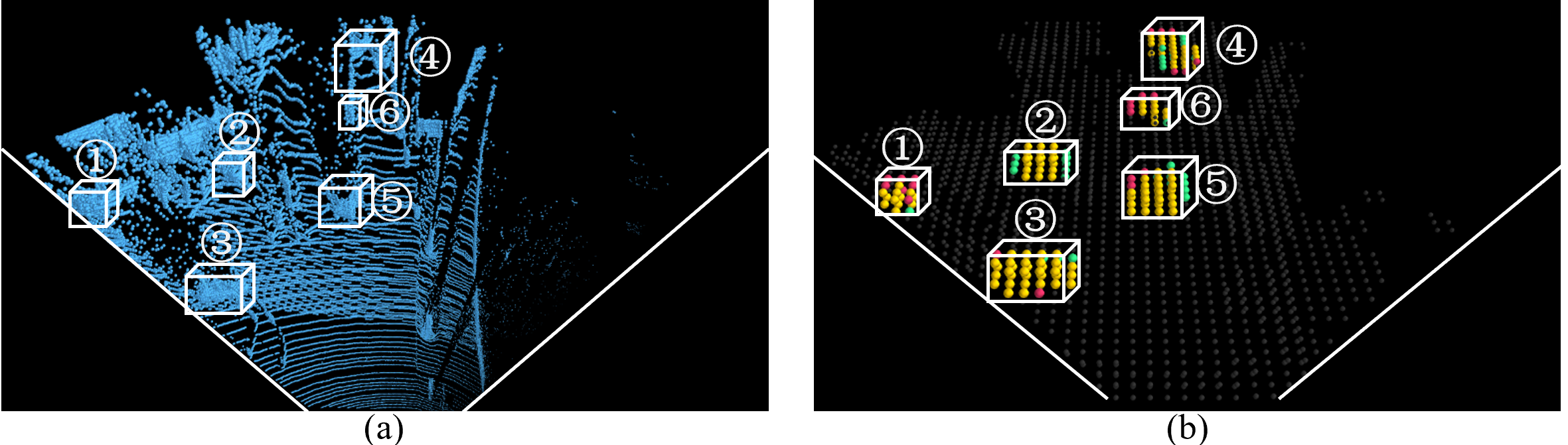}
    \caption{Channel compensation in the point selection process. (a) $\textbf{G}_\mathrm{V}$, (b) $\textbf{F}_\mathrm{4}$.}
    \label{figure6}
\vspace{-0.6cm} 
\end{figure}
Considering that each channel can capture a distinct attribute in the multi-channel feature, we use a channel compensation strategy to obtain more comprehensive candidate point sets. Specifically, we conduct the same point selection operation according to values in  $\textbf{F}_\mathrm{4,2}^m$. The set of selected points is denoted as $S_2^s = \left\{ \textbf{P}_{\mathrm{2},i}^s \mid \emph{i}=1,2,...,N_{2}^s \right\} $, where $N_{2}^s$ denotes the number of points selected by $\textbf{F}_\mathrm{4,2}^m$. The set of selected points after channel compensation is denoted as $S_{1\cup 2}^s$, $S_{1\cup 2}^s=S_{1}^s\cup S_{2}^s$.  Fig. \ref{figure6} illustrates the channel compensation strategy. Fig. \ref{figure6} (a) shows the original point cloud. Fig. \ref{figure6} (b) visualizes the points in $\textbf{F}_\mathrm{4}$. Yellow points belong to the intersection of $S_{1}^s$ and $S_{2}^s$, green or red points belong either to $S_{1}^s$ or to $S_{2}^s$, and gray points are removed. The figure illustrates how the channel compensation strategy can effectively improve the integrity of the object area. 

Finally, we add a post-processing operation on $S_{1\cup 2}^s$ to compensate for potential information loss caused by suboptimal slope thresholds and unsatisfactory enhancement of object area. In the post-processing, we calculate the smallest Euclidean distance from every point in the set \emph{S} to every point in the set $S_{1\cup 2}^s$. If this distance is smaller than the post-processing distance threshold $d_p$, the point is selected. Otherwise, the point is removed. Finally, all selected points and their corresponding features form the output of the FGC bock, denoted as $\textbf{F}_\mathrm{4}^f$.

\subsubsection{Accuracy-Friendly Feature Encoder}
In above mentioned Transmission-Friendly Feature Encoder, the introduction of MTE enhances the features of the object area and provides the condition for conducting slope-based fine-grained geometry compression. However, the high cost of self-attention calculation requires the maxpooling to reduce the spatial resolution of $\textbf{F}_\mathrm{4}$ and thus reduce the requirement on computing power. Moreover, the original detection model needs the multi-scale features $\textbf{F}_\mathrm{3}$ and $\textbf{F}_\mathrm{4}$ to conduct object prediction at the detection head. Compared with $\textbf{F}_\mathrm{3}$, the geometric loss of $\textbf{F}_\mathrm{4}^f$ comes from two aspects. First, the convolution operation conducted in Eq. (\ref{gongshi8}) makes the resolution of $\textbf{F}_\mathrm{4}$ become 1/2 of $\textbf{F}_\mathrm{3}$. Second, the maxpooling operation conducted in Eq. (\ref{gongshi9}) reduces the resolution of $\textbf{F}_\mathrm{4}^f$ to half that of $\textbf{F}_\mathrm{4}$. These geometric losses determine that the T-FFC mode will inevitably lead to a degradation in detection accuracy. Although the channel compression operation is also conducted to obtain compact $\textbf{F}_\mathrm{4}^f$, the experimental results show that channel compression does not result in significant accuracy loss. Based on these observations, we design the A-FFC mode to transmit more features and avoid a decline in detection accuracy. 

Specifically, in the A-FFC mode, we design three feature compression branches. The first branch directly uses the above-mentioned Transmission-Friendly Feature Encoder to compress and enhance $\textbf{F}_\mathrm{4}$ to obtain an approximate location of the object area. It is worth mentioning that the maxpooling kernel size is changed to 3×3×3 in the A-FFC mode. The experimental results with different kernel sizes are given in the supplementary material. The other two branches are designed to conduct spatial and channel compressions on $\textbf{F}_\mathrm{3}$ and $\textbf{F}_\mathrm{4}$ under the guidance of $\textbf{F}_\mathrm{4}^f$ to obtain near geometric-lossless compressed feature representation of the object area. Channel compression is implemented through two VirConv blocks, while spatial compression is implemented using the backward mapping operation. The backward mapping operation takes $\textbf{F}_\mathrm{4}^f$ as guidance to look for the positional corresponding points in $\textbf{F}_\mathrm{3}$ and $\textbf{F}_\mathrm{4}$. The other points are removed since they are related to non-object areas. The above processes can be formulated as
\begin{equation}
\label{gongshi11}
\textbf{F}_3^a=\mathcal{R}\left(\mathcal{B}\left(\mathcal{C}_8\left(\mathcal{C}_7\left(\textbf{F}_3\right)\right),\textbf{F}_4^f\right),\textbf{F}_3\right),
\end{equation}
\begin{equation}
\label{gongshi12}
\textbf{F}_4^a=\mathcal{R}\left(\mathcal{B}\left(\mathcal{C}_{10}\left(\mathcal{C}_9\left(\textbf{F}_4\right)\right),\textbf{F}_4^f\right),\textbf{F}_4\right),
\end{equation}
where $\textbf{F}_3^a$ and $\textbf{F}_4^a$ denote the newly added compressed features in the A-FFC mode, $\mathcal{B}$ denotes the backward mapping operation and $\mathcal{R}$ denotes the background removal operation. The number of output channels of $\mathcal{C}_7$ and $\mathcal{C}_9$ is set to 16, while the number of output channels of $\mathcal{C}_8$ and $\mathcal{C}_{10}$ is set to 2.
\begin{figure}
\centering
    \setlength{\abovecaptionskip}{0cm}
    \includegraphics[width=0.824\linewidth]{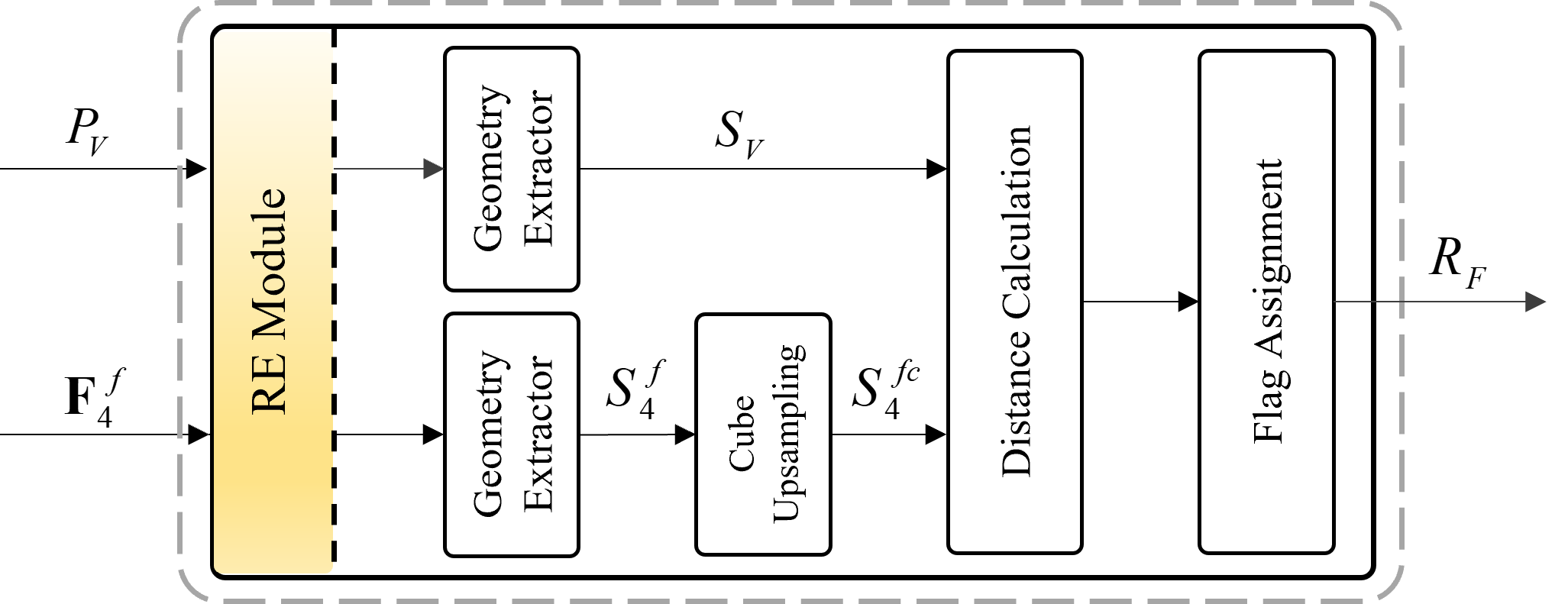}
    \caption{Residual extraction module.}
    \label{figure7}
\vspace{-0.6cm} 
\end{figure}
\subsection{Entropy Coding I and Entropy Decoding I}
Considering that the compressed sparse tensors have geometry and attribute information, we adopt the encoder of Oct-Attention \cite{ref69} and CABAC (i.e., $\mathcal{E}_{O}$ and $\mathcal{E}_{C}$) to achieve the transformation from geometry and attribute information to binary bitstreams, which can be formulated as
\begin{equation}
\label{gongshi13}
B_G= \mathcal{E}_{O}\left(\textbf{G}_E\right),
\end{equation}
\begin{equation}
\label{gongshi14}
B_A= \mathcal{E}_{C}\left(\textbf{A}_E\right),
\end{equation}
where $\textbf{G}_E$ and $\textbf{A}_E$ denote the geometry and attribute informationc obtained at the encoder side. For the A-FFC mode, they come from $\textbf{F}_\mathrm{4}^f$, $\textbf{F}_\mathrm{3}^a$ and $\textbf{F}_\mathrm{4}^a$. For the T-FFC mode, they only come from $\textbf{F}_\mathrm{4}^f$. $B_G$ and $B_A$ denote the binary bitstream of geometry and attribute information. 

At the cloud device, the decoder of Oct-Attention \cite{ref69} and CABAC (i.e., $\mathcal{D}_{O}$ and $\mathcal{D}_{C}$) are used to obtain the geometry and attribute information of the remaining points, which can be expressed as
\begin{equation}
\label{gongshi15}
\textbf{G}_D=\mathcal{D}_{O}\left(B_G\right),
\end{equation}
\begin{equation}
\label{gongshi16}
\textbf{A}_D=\mathcal{D}_{C}\left(B_A\right),
\end{equation}
where $\textbf{G}_D$ and $\textbf{A}_D$ denote the geometry and attribute information obtained at the decoder side used for the detection task. For the A-FFC mode, we can obtain $\hat{\textbf{F}}_\mathrm{4}^f$, $\hat{\textbf{F}}_\mathrm{3}^a$ and $\hat{\textbf{F}}_\mathrm{4}^a$. For the T-FFC mode, we can only obtain $\hat{\textbf{F}}_\mathrm{4}^f$.

\subsection{Feature Decoding}
\subsubsection{Transmission-Friendly Feature Decoder}

On the cloud device, the coordinates of points in $\hat{\textbf{F}}_\mathrm{4}^f$ are multiplied with the maxpooling ratio to reconstruct the spatial size. Then, the CE module is used to reconstruct the channel dimension. The channel reconstruction process can be formulated as
\begin{equation}
\label{gongshi17}
\textbf{F}_\mathrm{3d1}= \mathcal{C}_\mathrm{12}\left(\mathcal{C}_\mathrm{11}\left(\hat{\textbf{F}}_\mathrm{4}^f\right)\right),
\end{equation}
where the number of output channels of $\mathcal{C}_\mathrm{11}$ and $\mathcal{C}_\mathrm{12}$ is set to 16 and 64, respectively. 

After that, $\textbf{F}_\mathrm{3d1}$ is fed to the spatial upsampling module which consists of the transpose convolution layer and submanifold sparse convolution layer to obtain reconstructed features $\textbf{F}_\mathrm{3d2}$ and $\textbf{F}_\mathrm{3d3}$. The transpose convolution layer (i.e., $\mathcal{C}_\mathrm{1}^T$ and $\mathcal{C}_\mathrm{2}^T$) is used for spatial upsampling and the submanifold sparse convolution layer (i.e., $\mathcal{C}_\mathrm{1}^S$ and $\mathcal{C}_\mathrm{2}^S$) is used for feature enhancement. The above operations can be formulated as 
\begin{equation}
\label{gongshi18}
\textbf{F}_\mathrm{3d2}= \mathcal{C}_\mathrm{1}^S\left(\mathcal{C}_\mathrm{1}^T\left(\textbf{F}_\mathrm{3d1}\right)\right),
\end{equation}
\begin{equation}
\label{gongshi19}
\textbf{F}_\mathrm{3d3}= \mathcal{C}_\mathrm{2}^S\left(\mathcal{C}_\mathrm{2}^T\left(\textbf{F}_\mathrm{3d2}\right)\right),
\end{equation}
where the number of output channels, kernel sizes, stride, padding and dilation of $\mathcal{C}_\mathrm{1}^T$ and $\mathcal{C}_\mathrm{2}^T$ are set to 64, 3×3×3, 1, 2 and 2. The number of output channels and kernel sizes of $\mathcal{C}_\mathrm{1}^S$ and $\mathcal{C}_\mathrm{2}^S$ are set to 64 and 3×3×3. 
\begin{figure}
    \centering
    \setlength{\abovecaptionskip}{0cm}
    \includegraphics[width=1\linewidth]{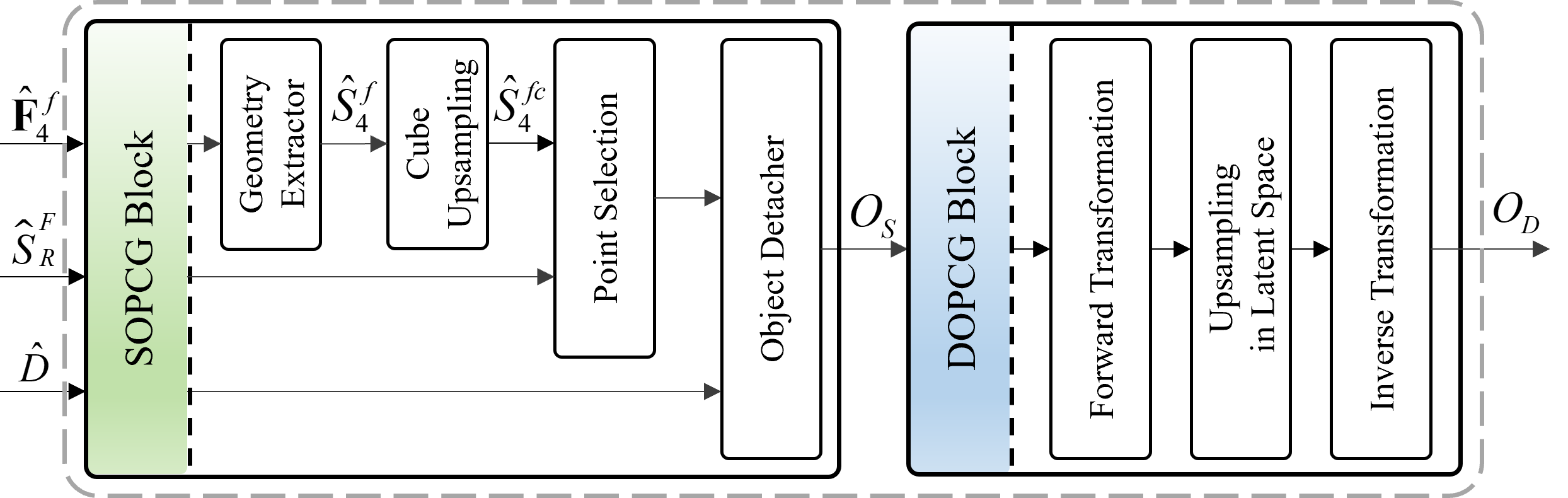}
    \caption{3D object reconstruction module.}
    \label{figure8}
\vspace{-0.6cm} 
\end{figure}
\subsubsection{Accuracy-Friendly Feature Decoder}

In this situation, we conduct the same channel expansion and scale reconstruction operations as the Accuracy-Friendly Feature Decoder on $\hat{\textbf{F}}_\mathrm{4}^f$ to obtain  $\textbf{F}_\mathrm{3d1}$. Additionally, $\hat{\textbf{F}}_\mathrm{3}^a$ and $\hat{\textbf{F}}_\mathrm{4}^a$ are fed to channel expansion modules to obtain $\textbf{F}_\mathrm{3d2}$ and $\textbf{F}_\mathrm{3d3}$. These processes can be formulated as
\begin{equation}
\label{gongshi20}
\textbf{F}_\mathrm{3d2}= \mathcal{C}_\mathrm{14}\left(\mathcal{C}_\mathrm{13}\left(\hat{\textbf{F}}_\mathrm{3}^a\right)\right),
\end{equation}
\begin{equation}
\label{gongshi21}
\textbf{F}_\mathrm{3d3}= \mathcal{C}_\mathrm{16}\left(\mathcal{C}_\mathrm{15}\left(\hat{\textbf{F}}_\mathrm{4}^a\right)\right),
\end{equation}
where the number of output channels of $\mathcal{C}_\mathrm{13}$, $\mathcal{C}_\mathrm{14}$, $\mathcal{C}_\mathrm{15}$, and $\mathcal{C}_\mathrm{16}$ are set to 16, 64, 16, and 64, respectively.

\subsection{Feature Analysis}
The stage two of VirConv-L is made up of the RPN and Detection Head. The RPN (i.e., $\mathcal{P}$) takes $\textbf{F}_\mathrm{3d1}$ as input and generates region proposals. Then, the region proposals, $\textbf{F}_\mathrm{3d2}$ and $\textbf{F}_\mathrm{3d3}$ are sent to the detection head (i.e., $\mathcal{H}$) to obtain predicted results $\hat{{D}}$. These processes can be formulated as
\begin{equation}
\label{gongshi22}
\hat{{D}} = \mathcal{H}\left(\textbf{F}_\mathrm{3d2},\textbf{F}_\mathrm{3d3},\mathcal{P}\left(\textbf{F}_\mathrm{3d1}\right)\right).
\end{equation}
\subsection{Residual Extraction}
Considering that $\textbf{F}_\mathrm{4}^f$ is the commonly transmitted data in the both two modes, the reconstruction model is designed on it to improve its applicability. However, if we directly use the $\textbf{F}_\mathrm{4}^f$ for reconstruction tasks, there will be the following problems: 1) The convolution and maxpooling operations conducted in Eq. (\ref{gongshi8}) and Eq. (\ref{gongshi9}) make the resolution of features low, which is unamiable to the fine-grained reconstruction task. 2) The 3D object detection task aims to describe the position and box size of interested objects. Therefore, the feature extracted in the detection task can only reflect the contour of the target while the local detail information is missing. 3) The feature extraction operations conducted on true and virtual points cause feature diffusion \cite{ref21}. The points located inside the objects mislead the reconstruction process. Therefore, we design the RE module to assist the reconstruction task and obtain better results. The framework of RE module is shown in Fig. \ref{figure7}. First, we conduct the geometry extractor operation on $\textbf{F}_\mathrm{4}^f$ and $P_V$ to obtain their coordinate point set $S_\mathrm{4}^f$ and $S_V$. Then the cube upsampling operation (i.e., $\mathcal{U}_C$) is conducted on $S_\mathrm{4}^f$ to achieve the geometric extension and generate the candidate point set $S_\mathrm{4}^{fc}$. The data volume of the extracted residual information can be adjusted by changing the scope and density parameter (i.e., $U_C^s$ and $U_C^d$) of the upsampling operation. This upsampling process is formulated as
\begin{equation}
\vspace{-0.15cm} 
\label{gongshi23}
{S}_\mathrm{4}^\emph{fc} = \mathcal{U}_C\left({S}_\mathrm{4}^\emph{f},U_C^s,U_C^d\right). 
\end{equation}

After that, the distance calculation and flag assignment operations are conducted to pick out the points belong to the surface of objects. Specifically, we first calculate the minimum Euclidean distance from every point in the set $S_\mathrm{4}^{fc}$ to every point in the set $S_V$.  If the calculated distance is smaller than the surface judgment distance threshold $d_s$, we think the point belongs to the surface of the object, thus it is selected and assigned with flag 1. Otherwise, it is removed and assigned with flag 0. To reduce the data volume of residual information, we transmit the flag set of points in $S_\mathrm{4}^{fc}$, denoted as $R_F$. 

\subsection{Entropy Coding II and Entropy Decoding II}
Considering that the flag 0 accounts for a large proportion in $R_F$, we only encode the position of flag 1 in the residual-related entropy encoder. Specifically, we first find the absolute position of flag 1 to obtain the sequence to be encoded, denoted as $S_{\emph{R}}^F$. Then, we encode $S_{\emph{R}}^F$ using the differential encoding algorithm to narrow the range of data values. Finally, the CABAC encoder is adopted to transform the relative position sequence to binary bitstream. The encoding process of $S_{\emph{R}}^F$ can be formulated as
\begin{equation}
\label{gongshi24}
{B}_\emph{R} = \mathcal{E}_{C}\left(\mathcal{E}_{D}\left({S}_\emph{R}^F\right)\right),
\end{equation}
where $\mathcal{E}_{D}$ denotes the encoder of differential encoding.

At the cloud device, the decoders of CABAC and differential encoding (i.e., $\mathcal{D}_{C}$ and $\mathcal{D}_{D}$) are used to reconstruct the residual, which can be formulated as
\begin{equation}
\label{gongshi25}
\hat{S}_\emph{R}^F = \mathcal{D}_{C}\left(\mathcal{D}_{D}\left({B}_\emph{R}\right)\right).
\end{equation}
\vspace{-0.85cm}
\subsection{3D Reconstruction}
As shown in Fig. \ref{figure8}, the framework of the 3DOR module is made up of the Sparse Object Point Cloud Generation (SOPCG) block and Dense Object Point Cloud Generation (DOPCG) block. A detailed description of these blocks is given below.
\subsubsection{SOPCG Block}First, we conduct the geometry extraction operation on $\hat{\textbf{F}}_\mathrm{4}^{f}$ to obtain the position set of points $\hat{\emph{S}}_\mathrm{4}^{f}$. Then, the same cube upsampling operation as in Eq. (\ref{gongshi23}) is conducted on $\hat{\emph{S}}_\mathrm{4}^{f}$ to achieve geometric extension and obtain $\hat{\emph{S}}_\mathrm{4}^{fc}$. Subsequently, $\hat{\emph{S}}_R^F$ is used for filtering out the points belonging to the surface of the object. After that, $\hat{\emph{D}}$ and filtered points are fed to the object detacher to obtain sparse object point clouds ${O}_\emph{S}$. The diversity of distance and occlusion of objects makes the point number in different objects various. We conduct a patch partition operation on objects to obtain training patches with the same size. The smaller the patch size is, the more objects can be reconstructed. However, the disadvantage is that a smaller patch size brings more challenges in the reconstruction process. Moreover, the larger patch size can better describe the structure of an object but there will be more objects that can’t be reconstructed. In this paper, the patch size is set to 30 and the up ratio is set to 4. The above processes can be formulated as
\begin{equation}
\label{gongshi26}
{O}_\emph{S} = \mathcal{O}\left(\mathcal{S}\left(\mathcal{U}_C\left(\hat{S}_4^f\right),\hat{S}_{R}^{F}\right),\hat{D}\right),
\end{equation}
where $\mathcal{S}$ denotes the point selection operation, $\mathcal{O}$ denotes the object detacher operation.
\subsubsection{DOPCG Block} As shown in Fig. \ref{figure8}, the DOPCG block consists of the forward transformation, upsampling, and inverse transformation operations. A detailed description of the upsampling network can be found in \cite{ref61}. The reconstruction process can be formulated as
\begin{equation}
\label{gongshi27}
{O}_\emph{D} = \mathcal{F}_I\left(\mathcal{U}_L\left(\mathcal{F}_F\left({O}_\emph{S}\right)\right)\right),
\end{equation}
where $\mathcal{F}_F$ denotes the transformation from point representation to latent representation, $\mathcal{F}_I$ denotes the transformation from latent representation to point representation, $\mathcal{U}_L$ denotes the upsample operation conducted in latent space. 

\subsection{Loss Function}

\subsubsection{Object Detection} The detection loss consists of the loss from RPN and the loss from the detection head. The detailed description of these losses can be found in \cite{ref9}.
\subsubsection{Object Reconstruction}In the reconstruction task, we adopt the reconstruction loss and prior loss to enhance the surface similarity and optimize the flow module. The detailed description of these losses can be found in \cite{ref61}.

\section{Experimental Results and Analysis}
\label{Sec-V}
In this section, we first introduce the benchmark datasets and experimental protocols. Then, we examine detection performance as a function of bitrate. Next, we conduct an ablation study. Afterward, we compare our method with CTA methods. Finally, we analyse the reconstruction results. 

\subsection{Datasets and Experimental Protocols}
\subsubsection{Datasets} We chose the commonly used KITTI dataset to evaluate the performance of the proposed method. For the multimodal 3D object detection task, 7481 and 7518 pairs of point cloud and image data were used for training and testing the performance of various detection methods. Similar to \cite{ref9}, the training set was further divided into training and validation subsets. The former consisted of 3712 pairs of multimodal data, and the latter consisted of 3769 pairs of multimodal data. Moreover, the objects in the KITTI dataset were divided into three categories (easy, moderate, and hard) in terms of detection difficulty. 
\subsubsection{Evaluation Criteria} We used the Average Precision (AP) for recall values less than or equal to 40\% to measure the accuracy of results. The Intersection over Union (IoU) threshold was set to 0.7. We reported the Car BEV AP and Car 3D AP to reflect detection accuracy in BEV and 3D view, respectively. We used the Chamfer Distance (CD) to measure the similarity between reconstructed objects and ground truth.

\subsection{R-P Curves of Object Detection Task}

To study the effect of the data volume on the detection task, we give the corresponding Rate-Performance (R-P) curve (Fig. \ref{figure9} and Fig. \ref{figure10}). The quantitative results are given in the supplementary material. Considering that both reducing $\emph{k}_\mathrm{TH}$ or increasing $\emph{d}_\emph{p}$ in the FGC module can improve the accuracy of the object detection task, we present two groups of results. 
\begin{figure}
    \centering
    \setlength{\abovecaptionskip}{0cm}
    \includegraphics[width=1\linewidth]{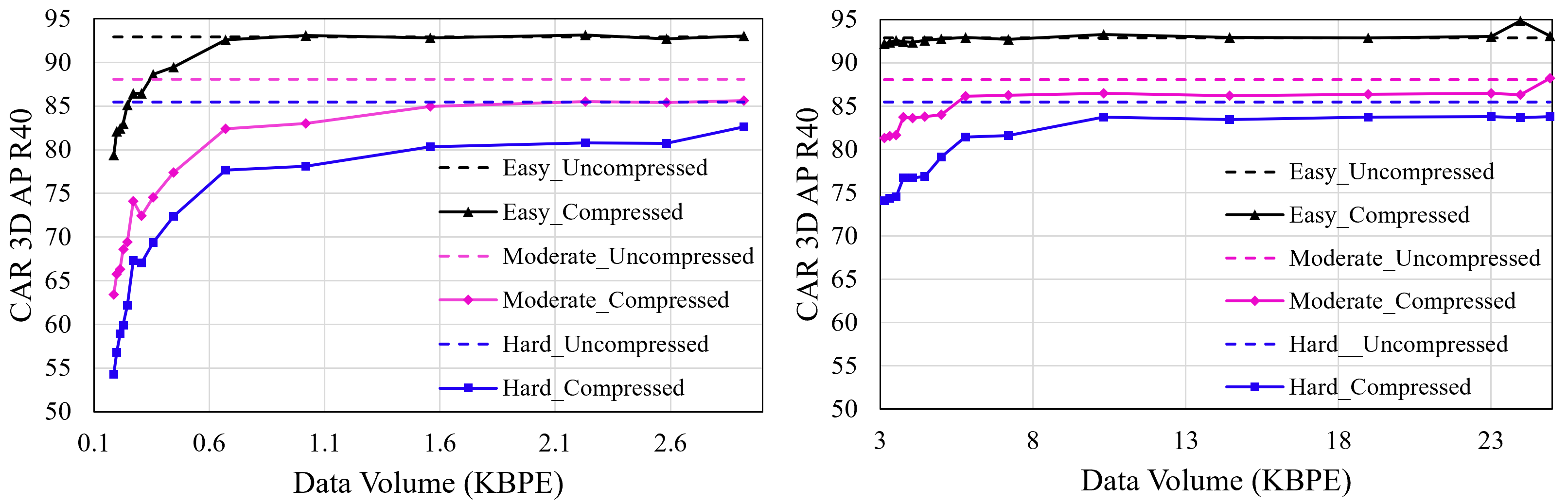}
    \caption{Slope threshold-changing based R-P curves of the detection task of A-FFC mode (left) and T-FFC mode (right).}
    \label{figure9}
\vspace{-0.6cm}
\end{figure}
In Fig. \ref{figure9}, $\emph{d}_\emph{p}$ in all testing processes was set to 0 and $\emph{k}_\mathrm{TH}$ was slowly changed to obtain various data volumes. As the data volume increased, the performance of easy objects saturates first, followed by moderate targets, and last by hard objects. Under the same $\emph{k}_\mathrm{TH}$, the A-FFC mode has larger data volume and better performance than T-FFC mode. The T-FFC mode achieves near lossless detection performance for easy objects when data volume is about 0.7 KiloBytes Per Element (KBPE). However, there were still performance drops for moderate and hard objects even when the data volume was close to 3 KBPE. Compared with the original VirConv-L model, the A-FFC achieved near-lossless detection performance for easy and moderate objects when data volume reached 25 KBPE. However, the A-FFC mode was unable to achieve lossless detection for hard objects. Fortunately, the introduced post-processing operation effectively addressed this issue. In Fig. \ref{figure10}, $\emph{k}_\mathrm{TH}$ was set to 0.02 and $\emph{d}_\emph{p}$ was changed from 0 to 9. As $\emph{d}_\emph{p}$ increased, the performance changes of A-FFC mode are slower than the performance changes of T-FFC mode. The reason may be that the maxpooling kernel size in A-FFC mode was 3×3×3, while it was 2×2×2 in T-FFC mode. Compared with $\emph{k}_\mathrm{TH}$ changing-based method, $\emph{d}_\emph{p}$ changing-based method achieved better performance with less data volume.

\subsection{Ablation Study}
In this subsection, we conduct a series of experiments to show the effectiveness of the proposed modules. The results are shown in Table \ref{Table IV} and Table \ref{Table V}. It is worth mentioning that the performance of schemes without maxpooling is not provided since they are computationally expensive. Moreover, the channel compression operation of basic feature in the CGC module cannot be removed either as directly applying a transformer enhancement operation on a 64-channel feature is infeasible. Although removing the transformer enhancement operation does not significantly harm detection performance, the numerical distinction between target and non-target regions in the features becomes less apparent, which causes the slope-based spatial downsampling strategy to fail in this situation. Besides, introducing convolution enhancement at the decoder side improved performance in T-FFC mode. Finally, we conducted an ablation study on channel compression of newly added features (i.e., $\textbf{F}_\mathrm{3}$ and $\textbf{F}_\mathrm{4}$) in A-FFC mode. The results indicate that channel compression of $\textbf{F}_\mathrm{3}$ and $\textbf{F}_\mathrm{4}$ did not lead to a significant degradation in the detection task while achieving efficient feature compression.

\begin{table*}[]
\centering
\belowrulesep=0pt
\aboverulesep=0pt
\setlength{\abovecaptionskip}{-0.1cm}
\caption{ABLATION STUDY OF THE T-FFC MODE. TOP THREE RESULTS ARE HIGHLIGHTED IN RED, GREEN, AND BLUE, RESPECTIVELY.}
\label{Table IV}
\begin{tabular}{c|c|c|c|c|c|c|c|c|c|c}
\toprule
 &
   &
   &
   &
   &
  \multicolumn{3}{c|}{Car BEV AP (R40)} &
  \multicolumn{3}{c}{Car 3D AP (R40)} \\ \cline{6-11} 
\multirow{-2}{*}{Scheme} &
  \multirow{-2}{*}{Maxpooling} &
  \multirow{-2}{*}{\begin{tabular}[c]{@{}c@{}}Channel compression\\of basic feature\end{tabular}} &
  \multirow{-2}{*}{\begin{tabular}[c]{@{}c@{}}Transformer\\enhancement\end{tabular}} &
  \multirow{-2}{*}{\begin{tabular}[c]{@{}c@{}}Convolution\\enhancement\end{tabular}} &
  Easy &
  Moderate &
  Hard &
  Easy &
  Moderate &
  Hard \\ \midrule
1 &
  × &
  × &
  × &
  × &
  \cellcolor[HTML]{C6DAFC}96.0807 &
  \cellcolor[HTML]{B2D9C6}92.0428 &
  \cellcolor[HTML]{FAC7C3}91.4400 &
  \cellcolor[HTML]{C6DAFC}92.8923 &
  \cellcolor[HTML]{FAC7C3}88.0660 &
  \cellcolor[HTML]{FAC7C3}85.4861 \\
2 &
  \checkmark &
  \checkmark &
  \checkmark &
  \checkmark &
  \cellcolor[HTML]{FAC7C3}96.3852 &
  \cellcolor[HTML]{FAC7C3}92.1788 &
  \cellcolor[HTML]{B2D9C6}89.4281 &
  \cellcolor[HTML]{B2D9C6}92.9419 &
  \cellcolor[HTML]{B2D9C6}85.8362 &
  \cellcolor[HTML]{B2D9C6}82.8727 \\
3 &
  \checkmark &
  \checkmark &
  × &
  \checkmark &
  \cellcolor[HTML]{B2D9C6}96.3376 &
  \cellcolor[HTML]{C6DAFC}91.9414 &
  86.9797 &
  92.6566 &
  83.0740 &
  77.8811 \\
4 &
  \checkmark &
  \checkmark &
  \checkmark &
  × &
  93.7519 &
  91.8294 &
  \cellcolor[HTML]{C6DAFC}89.0105 &
  \cellcolor[HTML]{FAC7C3}93.0963 &
  \cellcolor[HTML]{C6DAFC}85.7235 &
  \cellcolor[HTML]{C6DAFC}82.5818 \\ \bottomrule
\end{tabular}
\vspace{-0.4cm} 
\end{table*}

\begin{table*}[]
\centering
\belowrulesep=0pt
\aboverulesep=0pt
\setlength{\abovecaptionskip}{-0.1cm}
\caption{ABLATION STUDY OF A-FFC MODE. TOP THREE RESULTS ARE HIGHLIGHTED IN RED, GREEN, AND BLUE, RESPECTIVELY.}
\resizebox{\textwidth}{!}{
\label{Table V}
\begin{tabular}{c|c|c|c|c|c|c|c|c|c|c}
\toprule
 &
   &
   &
   &
   &
  \multicolumn{3}{c|}{Car BEV AP (R40)} &
  \multicolumn{3}{c}{Car 3D AP (R40)} \\ \cline{6-11} 
\multirow{-2}{*}{Scheme} &
  \multirow{-2}{*}{Maxpooling} &
  \multirow{-2}{*}{\begin{tabular}[c]{@{}c@{}}Channel compression\\of basic feature\end{tabular}} &
  \multirow{-2}{*}{\begin{tabular}[c]{@{}c@{}}Channel compression of\\newly added features\end{tabular}} &
  \multirow{-2}{*}{\begin{tabular}[c]{@{}c@{}}Convolution\\enhancement\end{tabular}} &
  Easy &
  Moderate &
  Hard &
  Easy &
  Moderate &
  Hard \\ \hline
1 &
  × &
  × &
  × &
  × &
  \cellcolor[HTML]{B2D9C6}96.0807 &
  92.0428 &
  \cellcolor[HTML]{FAC7C3}91.4400 &
  \cellcolor[HTML]{C6DAFC}92.8923 &
  \cellcolor[HTML]{B2D9C6}88.0660 &
  \cellcolor[HTML]{C6DAFC}85.4861 \\
2 &
  \checkmark &
  \checkmark &
  \checkmark &
  \checkmark &
  96.0324 &
  \cellcolor[HTML]{C6DAFC}92.0626 &
  \cellcolor[HTML]{C6DAFC}89.4763 &
  \cellcolor[HTML]{B2D9C6}92.9744 &
  \cellcolor[HTML]{FAC7C3}88.1588 &
  \cellcolor[HTML]{B2D9C6}85.4900 \\
3 &
  \checkmark &
  \checkmark &
  × &
  \checkmark &
  \cellcolor[HTML]{C6DAFC}96.0583 &
  \cellcolor[HTML]{B2D9C6}92.2058 &
  \cellcolor[HTML]{B2D9C6}89.6003 &
  \cellcolor[HTML]{FAC7C3}92.9756 &
  \cellcolor[HTML]{C6DAFC}88.0455 &
  \cellcolor[HTML]{FAC7C3}85.5285 \\
4 &
  \checkmark &
  \checkmark &
  \checkmark &
  × &
  \cellcolor[HTML]{FAC7C3}96.3820 &
  \cellcolor[HTML]{FAC7C3}92.5253 &
  87.5647 &
  92.7514 &
  86.0711 &
  81.1472 \\ \bottomrule
\end{tabular}}
\vspace{-0.4cm}
\end{table*}

\begin{table*}[]
\centering
\belowrulesep=0pt
\aboverulesep=0pt
\setlength{\abovecaptionskip}{-0.1cm}
\caption{PERFORMANCE OF RECONSTRUCTION TASK.}
\resizebox{\textwidth}{!}{
\label{Table XI}
\begin{tabular}{c|ccccc|cc|ccccc}
\toprule
\multirow{3}{*}{Scheme} &
  \multicolumn{5}{c|}{Parameters} &
  \multicolumn{2}{c|}{Data Volume (MB)} &
  \multicolumn{5}{c}{Metric} \\ \cline{2-13} 
 &
  \multicolumn{1}{c|}{\multirow{2}{*}{$\emph{k}_\mathrm{TH}$}} &
  \multicolumn{1}{c|}{\multirow{2}{*}{$\emph{d}_\emph{p}$}} &
  \multicolumn{1}{c|}{\multirow{2}{*}{$U_c^s$}} &
  \multicolumn{1}{c|}{\multirow{2}{*}{$U_c^d$}} &
  \multirow{2}{*}{$\emph{d}_\emph{s}$} &
  \multicolumn{1}{c|}{\multirow{2}{*}{\begin{tabular}[c]{@{}c@{}}Compressed basic\\ feature volume\end{tabular}}} &
  \multirow{2}{*}{\begin{tabular}[c]{@{}c@{}}Compressed\\ residual volume\end{tabular}} &
  \multicolumn{3}{c|}{Car 3D AP (R40)} &
  \multicolumn{1}{c|}{\multirow{2}{*}{\begin{tabular}[c]{@{}c@{}}Number\\ of object\end{tabular}}} &
  \multirow{2}{*}{\begin{tabular}[c]{@{}c@{}}Chamfer\\ Distance\end{tabular}} \\ \cline{9-11}
 &
  \multicolumn{1}{c|}{} &
  \multicolumn{1}{c|}{} &
  \multicolumn{1}{c|}{} &
  \multicolumn{1}{c|}{} &
   &
  \multicolumn{1}{c|}{} &
   &
  \multicolumn{1}{c|}{Easy} &
  \multicolumn{1}{c|}{Moderate} &
  \multicolumn{1}{c|}{Hard} &
  \multicolumn{1}{c|}{} &
   \\ \hline
1 &
  \multicolumn{1}{c|}{0.02} &
  \multicolumn{1}{c|}{2} &
  \multicolumn{1}{c|}{2} &
  \multicolumn{1}{c|}{1} &
  0.4 &
  \multicolumn{1}{c|}{4.1125} &
  4.9790 &
  \multicolumn{1}{c|}{92.8944} &
  \multicolumn{1}{c|}{83.1905} &
  \multicolumn{1}{c|}{80.2852} &
  \multicolumn{1}{c|}{6285} &
  0.012967 \\
2 &
  \multicolumn{1}{c|}{0.02} &
  \multicolumn{1}{c|}{4} &
  \multicolumn{1}{c|}{2} &
  \multicolumn{1}{c|}{1} &
  0.4 &
  \multicolumn{1}{c|}{7.2844} &
  7.1947 &
  \multicolumn{1}{c|}{92.8816} &
  \multicolumn{1}{c|}{85.1474} &
  \multicolumn{1}{c|}{80.6883} &
  \multicolumn{1}{c|}{9043} &
  0.011568 \\
3 &
  \multicolumn{1}{c|}{0.02} &
  \multicolumn{1}{c|}{6} &
  \multicolumn{1}{c|}{2} &
  \multicolumn{1}{c|}{1} &
  0.4 &
  \multicolumn{1}{c|}{11.0656} &
  9.4184 &
  \multicolumn{1}{c|}{92.9488} &
  \multicolumn{1}{c|}{85.6314} &
  \multicolumn{1}{c|}{82.6244} &
  \multicolumn{1}{c|}{13256} &
  0.009470 \\
4 &
  \multicolumn{1}{c|}{0.02} &
  \multicolumn{1}{c|}{4} &
  \multicolumn{1}{c|}{2} &
  \multicolumn{1}{c|}{2} &
  0.2 &
  \multicolumn{1}{c|}{7.2844} &
  15.0612 &
  \multicolumn{1}{c|}{92.8816} &
  \multicolumn{1}{c|}{85.1474} &
  \multicolumn{1}{c|}{80.6883} &
  \multicolumn{1}{c|}{14380} &
  0.007909 \\
5 &
  \multicolumn{1}{c|}{0.02} &
  \multicolumn{1}{c|}{6} &
  \multicolumn{1}{c|}{2} &
  \multicolumn{1}{c|}{2} &
  0.2 &
  \multicolumn{1}{c|}{11.0656} &
  21.0023 &
  \multicolumn{1}{c|}{92.9488} &
  \multicolumn{1}{c|}{85.6314} &
  \multicolumn{1}{c|}{82.6244} &
  \multicolumn{1}{c|}{14435} &
  0.007714 \\ \bottomrule
\end{tabular}}
\vspace{-0.4cm} 
\end{table*}

\subsection{Performance Comparison with CTA Model}
We present a performance comparison with the CTA model in Fig. \ref{figure11}. The quantitative results are given in the supplementary material. As for the multimodal input, the latest version of MPEG G-PCC test software (i.e., G-PCC-V23.0-rc2) and the state-of-the-art image coding standard VVC (i.e., VTM-22.2) were adopted to compress the point clouds and images respectively. The QPs of VVC were set to 51, 49, 47, 45, 43, and 41, and the compression configurations of G-PCC were set to R01, R02, R03, R04, R05, and R06. The results show that our method significantly outperformed the CTA method at low bitrates. 

\subsection{Performance of 3D Object Reconstruction Task}
\begin{figure}
    \centering
    \setlength{\abovecaptionskip}{0cm}
    \includegraphics[width=1\linewidth]{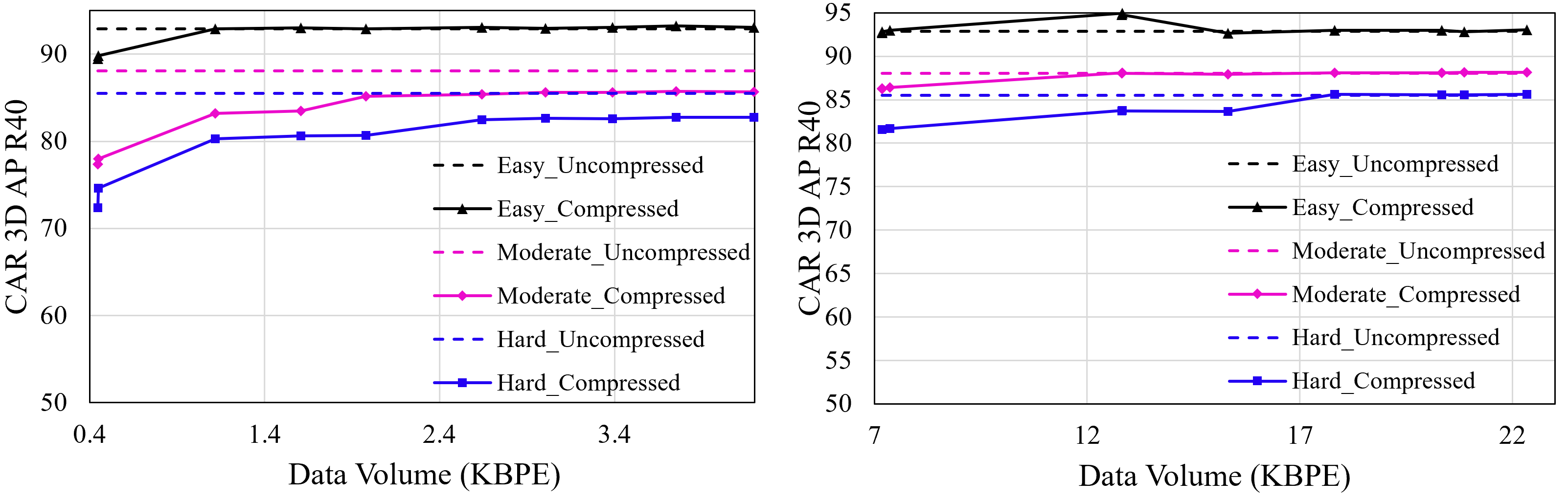}
    \caption{Radius-changing based R-P curve of the detection task of A-FFC mode (left) and T-FFC mode (right).}
    \label{figure10}
\vspace{-0.6cm} 
\end{figure}
Table \ref{Table XI} presents the quantitative results of the reconstruction task with different basic feature and residual sizes. With the increase of the data size of the residual, the performance of the reconstruction network initially rose rapidly, then stabilized as the data size of residual information gradually increased. Moreover, we provide the visualization results of the object reconstruction network in Fig. \ref{figure13}. The figure shows that as the data volume increased, the reconstructed objects became denser and the visual quality improved. However, there is still room for improvement in terms of details. This is because the occlusion and distance make the shape and size of objects used in this situation diverse. Therefore, the diversity of objects can promote the research of point cloud upsampling.
\begin{figure}[]
    \centering
    \setlength{\abovecaptionskip}{0cm}
    \includegraphics[width=1\linewidth]{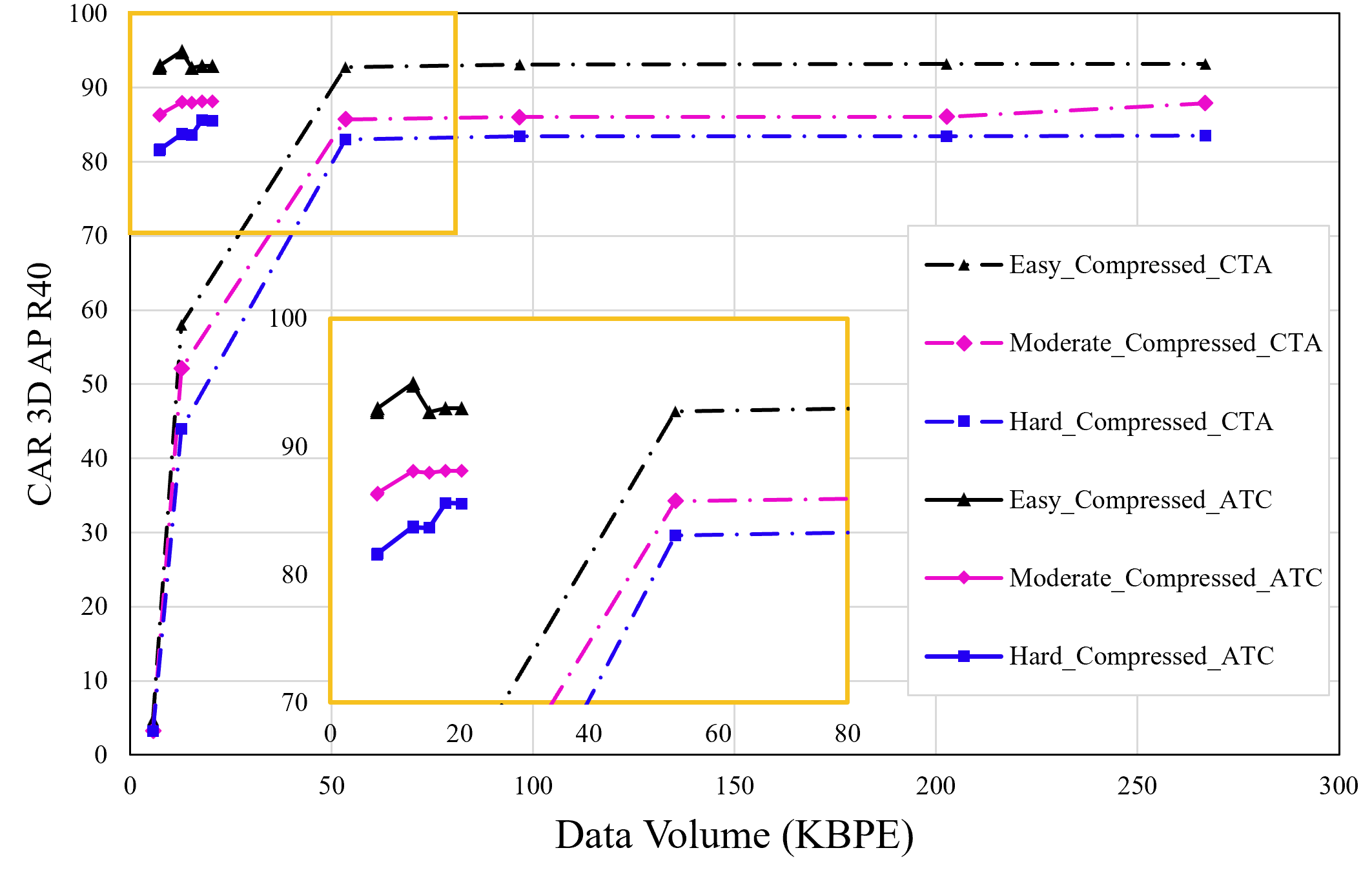}
    \caption{Performance comparison between CTA and ATC methods.}
    \label{figure11}
\vspace{-0.5cm} 
\end{figure}

\section{Conclusion}
\label{Sec-VI}
We proposed a feature compression method for multimodal 3D object detection. To address different application requirements, we designed two feature compression modes: Transmission-Friendly Feature Compression (T-FFC) mode and Accuracy-Friendly Feature Compression (A-FFC) mode. Experimental results demonstrate that the A-FFC mode achieves a feature compression ratio of 898 times with almost no loss in accuracy. Meanwhile, the T-FFC mode achieves a feature compression of 6061 times with a performance reduction of no more than 3\%. Additionally, we introduced an optional object reconstruction branch in this work. The reconstruction results provide an intuitive display of the shape, occlusion level, and other details of the detected objects. Overall, in a cloud-edge cooperation scenario, the proposed method significantly reduced data transmission and storage pressure with minimal to no loss in machine perception. Additionally, compared to transmitting raw data directly, transmitting features enhances data security.

In the future, we plan to enhance the proposed method in two key areas. First, we aim to design an efficient model pruning technique to reduce the computational complexity of edge devices. Second, we intend to develop an object reconstruction model with distortion correction capabilities to compensate for errors introduced by inaccurate virtual point estimation. 

\begin{figure}
    \centering
    \includegraphics[width=1\linewidth]{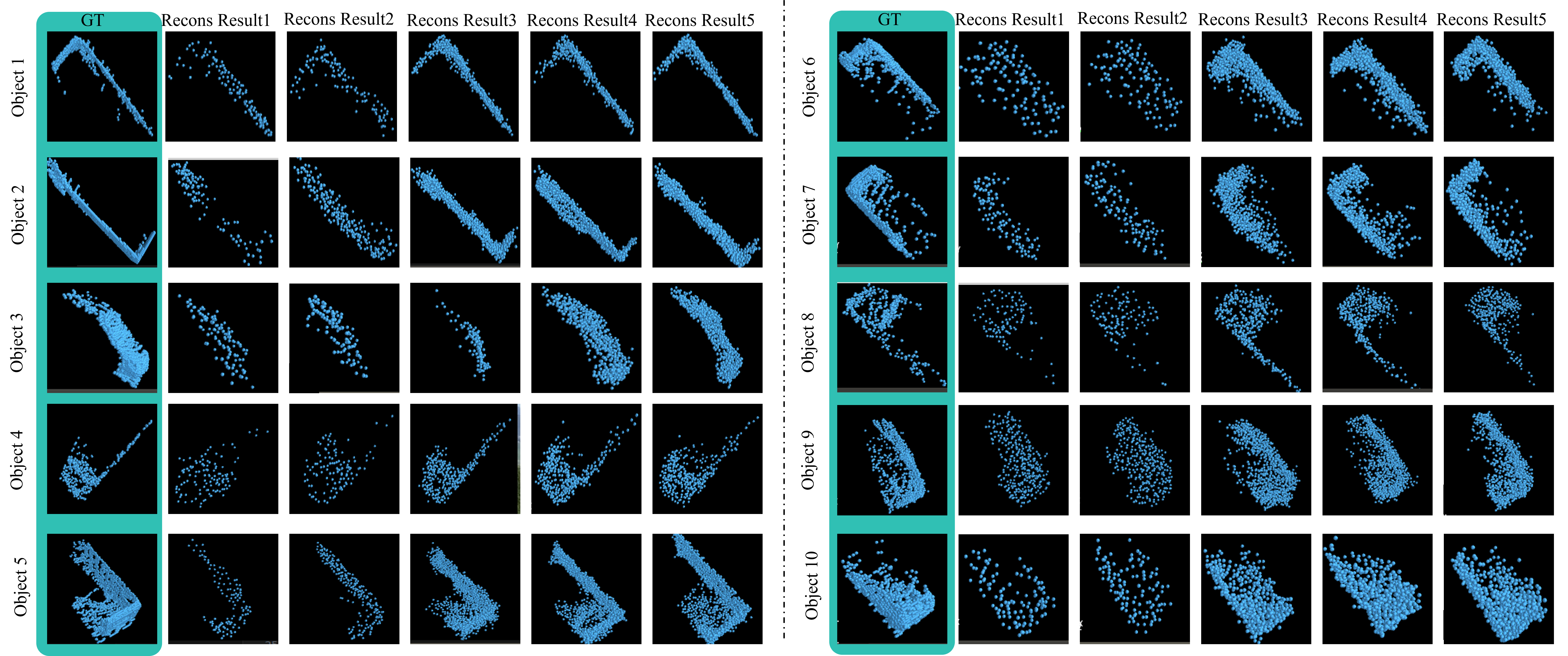}
    \caption{Visualization of the 3D object reconstruction results.}
    \label{figure13}
\end{figure}
\footnotesize
\bibliographystyle{IEEEbib}
\bibliography{240822_Main_Body}
\vspace{-0.5cm}
\begin{IEEEbiographynophoto}{Chongzhen Tian} received the B.S. degree and M.S. degree from Ningbo University, Ningbo, China, in 2020 and 2023, respectively. He is currently pursuing the Ph.D. degree at the Shandong University, Jinan, China. His research interests include point cloud compression and quality assessment.	
\end{IEEEbiographynophoto}
\vspace{-0.5cm}
\begin{IEEEbiographynophoto}{Zhengxin Li} received the B.S. degree from Shandong University, Weihai, China, in 2022. He is currently working toward the M.S. degree at Shandong University, Jinan, China. His current research interests include video/point cloud coding for machines.		
\end{IEEEbiographynophoto}
\vspace{-0.5cm}
\begin{IEEEbiographynophoto}{Hui Yuan} (Senior Member, IEEE) received the B.E. and Ph.D. degrees in telecommunication engineering from Xidian University, Xi’an, China, in 2006 and 2011, respectively. In April 2011, he joined Shandong University, Ji’nan, China, as a Lecturer (April 2011–December 2014), an Associate Professor (January 2015-August 2016), and a Professor (September 2016). His current research interests include 3D visual media coding, processing, and communication.	
\end{IEEEbiographynophoto}
\vspace{-0.5cm}
\begin{IEEEbiographynophoto}{Raouf Hamzaoui} (Senior Member, IEEE) received the M.Sc. degree in mathematics from the University of Montreal, Montreal, QC, Canada, in 1993, the Dr.rer.nat. degree from the University of Freiburg, Freiburg im Breisgau, Germany, in 1997, and the Habilitation degree in computer science from the University of Konstanz, Konstanz, Germany, in 2004. He was an Assistant Professor with the Department of Computer Science, University of Leipzig, Leipzig, Germany, and Department of Computer and Information Science, University of Konstanz.	In 2006, he joined De Montfort University, Leicester, U.K., where he is currently a Professor of media technology and the Head of the Signal Processing and Communications Systems Group, Institute of Engineering Sciences.
\end{IEEEbiographynophoto}
\vspace{-0.5cm}
\begin{IEEEbiographynophoto}{Liquan Shen} received the B.S. degree in automation control from Henan Polytechnic University, Jiaozuo, China, in 2001, and the M.E. and Ph.D. degrees in communication and information systems from Shanghai University, Shanghai, China, in 2005 and 2008, respectively. Since 2008, he has been with the Faculty of the School of Communication and Information Engineering, Shanghai University, where he is currently a Professor. His research interests include versatile video coding, perceptual coding, video codec optimization, 3-Dimensional Television, and video quality assessment.	
\end{IEEEbiographynophoto}
\vspace{-0.5cm}
\begin{IEEEbiographynophoto}{Sam Kwong} (Fellow, IEEE) is currently the Chair Professor of computational intelligence, and concurrently as Associate Vice-President (Strategic Research) of Lingnan University, Hong Kong. He is a Distinguished Scholar of evolutionary computation, artificial intelligence (AI) solutions, and image/video processing, with a strong record of scientific innovations and real-world impacts. 		
\end{IEEEbiographynophoto}

\newpage

\vfill

\end{document}